\documentclass[prd,aps,nofootinbib,floatfix,twocolumn,10 pt]{revtex4-1}

\usepackage{amssymb} \usepackage{amsmath,amssymb,graphicx,color,epsfig}
\usepackage{pstricks} \usepackage{float}

\begin{document}
\title{Inflationary study of non-Gaussianity using two-dimensional\\ geometrical measures of CMB temperature maps}
\author{M.  Junaid$^{1,2}$\footnote{mjunaid@ualberta.ca}
and D.  Pogosyan$^{1}$\footnote{pogosyan@ualberta.ca}}
\affiliation{$^1$Department of Physics, University of Alberta, Edmonton, Canada,\\
$^2$National Centre for Physics, Islamabad, Pakistan.}

\date{\today}

\begin{abstract}
In this work we effectively calculated the two-dimensional Minkowski functionals for cosmic microwave background (CMB) temperature maps generated by single field models of inflation with standard kinetic term. We started with a calculation of the bispectrum of initial perturbations and then calculated the two-dimensional configuration space cubic moments for temperature fluctuations. These cubic moments give rise to first order non-Gaussian correction terms to the Minkowski functionals. Thus, we developed a robust mechanism to predict the amount of non-Gaussianity generated by inflation in the CMB temperature maps using Minkowski functionals. 
\end{abstract}

\maketitle

\section{Introduction}\label{intro}
Among the most direct data sets for studies of the early inflationary stage
of our Universe evolution are the cosmic microwave background (CMB) temperature
and polarization maps that are becoming measured with high precision by ground and satellite based experiments \cite{planck4,wmap1,planck1,planck2,planck3,Planck15,boom,maxi,cbi,acbar}.
The primordial perturbations generated during inflation 
and reflected in CMB anisotropy are close to being Gaussian; however
small non-Gaussian contributions that reflect
specific features and peculiarities of inflationary models
are also in general expected to be present. 
The study of resulting non-Gaussian features in CMB maps
is promising for distinguish between these details of inflationary models. 

Inflation is the initial accelerated expansion of the early universe. An inflationary expansion of
more than $65$ $e$-folds is needed to explain the observed homogeneity and isotropy of the Universe.
It is predominantly accepted that inflation is driven by the potential energy of a scalar inflaton
field slowly rolling down the potential. Inflation also successfully describes the creation 
of small inhomogeneities, needed to seed observed structure in the Universe,
as having been generated by quantum fluctuations in the inflaton field. The quantum fluctuations
also got stretched and became imprinted on CMB maps and other observables on
cosmological scales. Thus, inflation is able to explain not only why the Universe is so 
homogeneous and isotropic but also the origin of the structures in the Universe
\cite{Guth1,Guth2,Linde3,mukhanov2,Bardeen2,starobinsky83,mukhanov}.
Alternatives to inflation have been proposed but no other scenario is as simple and elegant as
inflation produced by scalar field(s)~\cite{Linde1,Linde2,Kofman02}.

In single-field models of inflation, the generated initial inhomogeneities
are described via a single scalar adiabatic perturbation field $\zeta(\mathbf{x})$. 
The statistical properties of $\zeta(\mathbf{x})$ are the main observable signatures
to distinguish between different inflationary models.
We also know that if the perturbations are exactly Gaussian then all odd $n$-point
correlation functions vanish while all even $n$-point functions are related to
the two-point function. Thus, in momentum space the Gaussian field $\zeta(x)$
is completely described by the power spectrum $P_{\zeta}(k)$ given by
\begin{equation}
\left\langle{\zeta}_{\bold{k}}{\zeta}_{\bold{k}'}\right\rangle=
(2\pi)^3\delta^3(\bold{k}+\bold{k}')\frac{2\pi^2}{k^3}P_{\zeta}(k)\label{2ptf}.
\end{equation}
Inflation predicts that the scalar power spectrum is nearly flat
$P_\zeta(k) =A_s \left(\frac{k~}{k_*}\right)^{n_s-1}\approx A_s$, where $n_s$
is called the scalar spectral index\cite{mukhanov2,starobinsky83,Bardeen2}.
The fact that the observed value of scalar spectral index
$n_s=0.9603\pm0.0073$ \cite{planck2} is close to
but not exactly unity is considered as a strong evidence for the existence of inflation during
the early universe. However, many different kinds of inflationary
models can be made compatible with observations of the power spectrum.
Thus, the study of the non-Gaussian signatures is important to reduce the
degeneracy in inflationary models. Such signatures are contained in nontrivial
higher-order correlations starting with the cubic ones. These higher order
correlation functions also give us more insight about the physics of early universe. 

Similar to the power spectrum, for three-point correlations one can calculate the bispectrum
$B_{\zeta}(k_1,k_2,k_3)$ as a measure of the non-Gaussianity of the initial perturbations 
\begin{equation}
\left\langle\zeta_{\bold{k}_1}\zeta_{\bold{k}_2}\zeta_{\bold{k}_3}\right\rangle=
(2\pi)^3\delta^3(\bold{k}_1+\bold{k}_2+\bold{k}_3)B_{\zeta}(k_1,k_2,k_3)~.
\label{bis}
\end{equation}
The bispectrum carries much more information than the power spectrum as it contains three different length scales. 
It was shown by Maldacena that for basic single-field slow-roll inflation with a standard kinetic term the non-Gaussian effects
are small \cite{maldacena}. However, there are many models of inflation that give relatively large, potentially detectable, level of non-Gaussianity. Comparing non-Gaussian predictions with observations will help us
to constrain or rule out different inflationary models and give us more insight about the
physics of the early universe.

In this paper we will focus on the study of non-Gaussianity from inflation through real space
measures that describe geometrical and topological properties of 
CMB temperature maps viewed as a random field. Examples of standard measures of random fields are Minkowski functionals (MF), extrema statistics \cite{Bardeen3}
and also more novel
measures such as skeleton statistics \cite{Sousbie1,Pogosyan3}. Minkowski functionals are the most intuitive descriptors
of the properties of the excursion regions of the field above a certain threshold. Minkowski functionals have several mathematical properties that make them
special among other geometrical quantities. They are translationally and rotationally
invariant, additive, and have simple geometrical meanings. In \cite{hadwiger1957} it was shown that all global morphological properties  of any pattern in N-dimensional space that satisfy motional invariance and additivity can by fully characterized by $N + 1$ Minkowski Functionals. Indeed, for 2D field there are three functionals, which are simply the volume of the space where the field exceeds the threshold, the length of the boundary of the excursion set, and the Euler characteristic, $\chi_{2D}$ -- the number of separate connected regions with high field values minus the number of "holes" within them. In practice, we
are interested in statistical average Minkowski functionals as functions of threshold $\nu$ per unit volume of space.

It has been shown that for
mildly non-Gaussian field geometrical characteristics can be
expressed as a series of higher order moments of the perturbation field and
its derivatives~\cite{Matsubara1,PGP09,PPG11,Pogosyan1,Matsubara2,Pogosyan2}. 
In particular, the Euler characteristic density of excursion sets as a function of threshold
is given for a two-dimensional (2D) perturbation field up to first non-Gaussian correction by~\cite{Matsubara1,PGP09}
\begin{eqnarray}
\chi_\textsc{2D}(\nu) \approx \left(\frac{\sigma_1}{\sqrt{2} \sigma} \right)^2
\frac{e^{-\frac{\nu^2}{2}}}{(2\pi)^{3/2}} 
\Bigg[ H_1(\nu) +\sigma \left(\frac{
\left\langle\zeta^3 \right\rangle }{6\sigma^4} H_4(\nu) \right. \Bigg.\notag \\
\Bigg.
\left. -\frac{\left\langle\zeta^2\Delta\zeta \right\rangle}
{2\sigma^2\sigma_1^2} H_2(\nu)
- \frac{\left\langle(\nabla\zeta)^2\Delta\zeta \right\rangle}{\sigma_1^4}
\right) \Bigg].\quad
\label{genus}
\end{eqnarray}
In the above expression $\sigma^2=\left\langle\zeta^2\right\rangle$,
$\sigma_1^2=\left\langle(\nabla\zeta)^2\right\rangle$, $\nu$ is the threshold in units of $\sigma$ and
$H_i(\nu)$ are Hermite polynomials. The first term in the expansion denotes the Gaussian part that is proportional
to $H_1(\nu)$ while the terms that involve cubic moments of the field 
represent the first non-Gaussian correction in $\sigma$. On this example we see that real space geometrical characteristics
typically involve all hierarchy of moments. However in perturbation theory, higher order moments appear only in higher orders
of perturbative expansion, and the first non-Gaussian corrections involve only cubic moments of the field and its derivatives.

Minkowski functionals have been extensively used to characterize CMB maps since 
\cite{Gottetal1990}, even before they were introduced to cosmology on a formal basis in \cite{MeckeBuchertWagner1994}. They were measured in the first CMB maps by 
COBE satellite \cite{SchmaltzingGorski1998}, are WMAP data \cite{Eriksenetal2004} and
have been applied to the recent Planck CMB temperature maps \cite{Ducout1,Planck2013iso,Planck2015iso}. However, their direct use
to test predictions of the inflationary models was hindered by the difficulty to compute theoretical predictions from inflation for real space statistics, Minkowski functionals in particular.

We previously worked out the Minkowski functionals
for 3D perturbation field $\zeta$ at the end of inflation in single-field
models of inflation~\cite{Junaid}. 
In this paper we have developed a robust mechanism to compute theoretical
predictions for third-order
moments such as $\left\langle \zeta^3\right\rangle$, $\left\langle \zeta^2\Delta\zeta \right\rangle$ and
$\left\langle (\nabla\zeta)^2\Delta\zeta \right\rangle$ on 2D CMB maps that
are generated from these inflationary initial conditions.
This links the non-Gaussianity generated by inflation to the
geometrical observables such as Minkowski functionals in CMB maps.

This paper is organized into six sections. In Sec.~II, we review the
theoretical framework of inflationary cosmology whereafter we will describe
the calculation of the three-point correlation function in momentum space and
the calculation of different third-order moments in configuration space. In Sec.~III, 
we will present our numerical technique for the calculation of three-point
function and briefly discuss different single-field models of inflation with
some features in the inflationary potential. In Sec.~IV, we will present
the calculation of moments in configuration space while in
Sec.~V we present the geometrical Minkowski functionals.
In the last section we will summarize our results and conclude.


\section{Theoretical Framework}
The single-field models inflation driven by a scalar field
$\phi$ is described by the following action in units of ($M_{pl}^{-2} = 8\pi
G=1$, $c=\hbar=1$)
\begin{eqnarray} S = \int d^4x \sqrt{-g} 
\left(
\frac{1}{2} R - \frac{1}{2} g^{\mu\nu} \partial_{\mu}\phi \partial_{\nu}\phi -V(\phi)
\right)
\label{action}
\end{eqnarray}
where $V(\phi)$ is the potential for the
inflaton field. In Friedmann cosmology with homogeneous and isotropic
background, the Friedmann equation for scale factor and the Kline-Gordon equation
for inflaton field are given by
\begin{eqnarray}
H^2=\frac{1}{3}\left(\frac{1}{2}\dot{\phi}^2 +V(\phi) \right)\notag\\
\ddot{\phi}+3H\dot{\phi}+V_{,\phi}=0~.
\label{bg}
\end{eqnarray}
One can define the following slow-roll parameters\footnote{Our definition of
$\eta=\frac{\dot{\epsilon}}{\epsilon H}=2\epsilon-2\eta_\textsc{H}$ is commonly
used in the studies of non-Gaussianity whereas
$\eta_\textsc{H}=-\ddot{\phi}/(\dot{\phi}H)$ is the Hubble slow-roll parameter
used more commonly in the studies of inflation.} and the corresponding slow
roll conditions as
\begin{eqnarray}
\epsilon = -\frac{\dot{H}}{H}\ll1\text{, }
\eta = \frac{\dot{\epsilon}}{\epsilon H}\ll1~.
\end{eqnarray}
These slow-roll
conditions $\epsilon\ll1, \eta\ll1$ ensure that the inflaton field rolls slowly
down the potential and the Universe inflates for significantly long period.
These slow-roll parameters depend on potential of the inflaton field and the
model of inflation. For standard single-field inflation with quadratic
potential $V(\phi)=\frac{1}{2}m^2\phi^2$ these slow-roll parameters are of
order $O(0.01)$ for inflaton field values $\phi>10M_{pl}$. In the case of quadratic
inflation, to obtain 70 e-folds of inflation, one needs the initial inflaton field value to be $\phi_i\approx16.76\; M_{pl}$

\subsection{Calculation of Power Spectrum and Bispectrum}
In this section we will present the steps laid down by Maldacena to calculate
the two-point and the three-point correlation function of the scalar
perturbations \cite{maldacena}.
Firstly, one writes the action for the inflaton field given
in Eq.~\ref{action} using the Arnowitt-Deser-Misner(ADM) formalism. Secondly,
one expands the action to second order in perturbation theory for calculation
of the two-point function and to third-order for the calculation of three-point
function. Thirdly, one quantizes the perturbations and imposes canonical
commutation relations. Next, one can define the vacuum state by matching the
mode function to Minkowski vacuum when the mode is deep inside the horizon that
fixes the mode function completely. Following these steps one can find the
power spectrum and the bispectrum for scalar perturbations
\cite{maldacena,baumann1}.

In the ADM formalism the space-time is sliced into three-dimensional
hypersurfaces $\Sigma$, with three metric $g_{ij}$, at constant time. The line
element of the space-time is given by
\begin{eqnarray}
ds^2=-N^2dt^2+g_{ij}(dx^i+N^idt)(dx^j+N^jdt)
\end{eqnarray}
where $N$ and $N^i$ are lapse and shift functions. In single-field inflation, we only have one
physically independent scalar perturbation. Thus, we perturb the metric and
matter part of the action and use the gauge freedom to choose the comoving
gauge for the dynamical fields $\phi$ and $g_{ij}$
\begin{eqnarray}
\delta\phi = 0,& g_{ij}=a^2( e^{2\zeta}\delta_{ij}+t_{ij})~,
\label{gauge}
\end{eqnarray}
where $\zeta$ is the comoving curvature perturbation at constant density
hyper-surface $\delta\phi = 0$. In this gauge, the inflaton field is
unperturbed and all scalar degrees of freedom are parametrized by the metric
fluctuations $\zeta(t,x)$ while the tensor perturbations are parametrized by
$t_{ij}$, that is both traceless and orthogonal $\partial_it_{ij}=t^i_i=0$. The
conditions in Eq.~\ref{gauge} fixes the gauge completely at non zero momentum
\cite{maldacena}.  The shift and lapse functions are not dynamical variables in 
ADM formalism hence they can be derived from constraint equations in terms of $\zeta$. 
We study only scalar perturbations in this paper.

Linear perturbation results are obtained if one expands the action to second
order in perturbation field $\zeta(x)$
\begin{equation}
S_{(2)}= \int d^4x   a^3
\epsilon \left( \dot{\zeta}^2 - a^{-2}(\partial \zeta)^2 \right)
\end{equation}
which gives the following equation of motion for scalar perturbations
$v=z\zeta$ in Fourier space
\begin{equation}
v''_{\bold{k}} + \left(k^2
-\frac{z''}{z}\right)v_{\bold{k}} =0.
\label{Mukha}
\end{equation}
where $z=a\dot{\phi}/H$ and momentum $k$ is in reduced Planck units $M_{pl}$.
This is known as the Mukhanov equation for scalar perturbations\cite{mukhanov3}.
Now, one can calculate the power spectrum using Eq. \ref{2ptf} by calculating the two-point function of $\zeta(x)$
\begin{eqnarray}
P_{\zeta}(k)=\frac{k^3}{2\pi^2}|u_k|^2 
\label{PowerS}
\end{eqnarray}
where $u_k=v_k/z$ are the Fourier coefficients of $\zeta(x)$, the curvature
perturbations.

To obtain next order results in perturbation theory and to calculate
non-Gaussianity, one expands the action to third order in scalar perturbations
in the comoving gauge\cite{maldacena,DSeery}. After several integrations by parts
and dropping the total derivatives one finds the following third-order action
is often quoted in the literature\cite{maldacena,DSeery,Arroja,Horner1,DSeery2}
\begin{widetext}
\begin{eqnarray} S_{(3)} &=&
\int d^4x\bigg(a^3\epsilon(\epsilon-\eta)\zeta\dot{\zeta}^2
+a\epsilon^2\zeta(\partial\zeta)^2  -
\frac{a}{2}\epsilon\eta\zeta^2\partial^2\zeta -
2a\epsilon\dot{\zeta}(\partial_i\zeta)(\partial_i\chi)
+\frac{1}{2a}\epsilon\partial^2\chi(\partial_i\zeta)(\partial_i\chi)
+\frac{1}{4a}\epsilon(\partial^2\zeta)(\partial_i\chi)^2 \bigg. \label{S3}  \\
\bigg.&+& 2g(\zeta)\frac{\delta L}{\delta\zeta} \bigg),\quad
g(\zeta)=\zeta\dot{\zeta}/H+\frac{1}{4a^2H^2}\left[
-(\partial\zeta)^2+\partial^{-2}(\partial_i\partial_j(\partial_i\zeta\partial_j\zeta))
\right] + \frac{1}{4a^2H}\left[
-(\partial\zeta)(\partial\chi)+\partial^{-2}(\partial_i\partial_j(\partial_i\zeta\partial_j\chi))
\right]\notag
\end{eqnarray}
\end{widetext}
Here the last term can be eliminated with a field redefinition $\zeta \rightarrow \zeta_n+g(\zeta)$
because $g(\zeta)$ is only a function of derivatives of scalar perturbations
$\zeta(t,x)$ that vanish outside the horizon. The above third-order action is
an exact result without any slow-roll approximations; thus it is even valid for
models that deviate from slow-roll conditions. Another feature of this action
is that it contains only first two slow-roll parameters $\epsilon$ and $\eta$
while it is independent of derivative terms such as $\eta'$.

Finally, to calculate the three-point function in momentum space we move to the
interaction picture and write the Hamiltonian for the action in Eq.~\ref{action} as
\begin{equation}
H(\zeta) = H_0(\zeta) + H_{int}(\zeta)
\end{equation}
where $H_0$ is the quadratic part of the Hamiltonian while
$H_{int}$ represents all higher order terms in perturbation
theory\cite{maldacena}.
The three-point function is calculated using the in-in formalism in the interaction picture
using the third order action. Next, we quantize the perturbation field $\zeta(x)$ and
define the vacuum state. For this we expand the $\zeta(x)$ field into creation
and annihilation operators and use the commutation relations of the scalar field to get the
following result
\begin{eqnarray}
\left\langle
\zeta_{\bold{k}_1}\zeta_{\bold{k}_2}\zeta_{\bold{k}_3} \right\rangle = i
(2\pi)^3\int_{-\infty}^{\tau_{end}} d\tau \left(-2a^2\epsilon^2 u^*_1 u'^*_2
u'^*_3 \frac{\bold{k}_1.\bold{k}_2}{k_2^2} \right. \notag \\
\left.+2a^2\epsilon(\epsilon-\eta) u^*_1 u'^*_2 u'^*_3
-a^2\epsilon(2\epsilon\bold{k}_1.\bold{k}_2+\eta k_3^2) u^*_1 u^*_2 u^*_3
\right.\notag \\ + \frac{a^2}{2}\epsilon^3 u^*_1 u'^*_2 u'^*_3 k_1^2
\frac{\bold{k}_2.\bold{k}_3}{k_2^2 k_3^2 } +\frac{a^2}{2}\epsilon^3u^*_1 u'^*_2
u'^*_3 \frac{\bold{k}_1.\bold{k}_2}{k_2^2}+c.c.\notag \\ \bigg.+ \text{distinct
permutations}\bigg) \left.\prod_{i=1}^3 u_{i}(\tau_{end})\right.\delta^3(\sum_j
\bold{k}_j)\quad \label{3pt}
\end{eqnarray}
The choice of the vacuum is
specified by the choice of mode function $u_k$ selection. The above expression 
for the three-point function will be in used latter sections for the exact and
numerical calculation of three-point function in momentum space.

\subsection{Non-Gaussian observables}
The integral relation for three-point function given in Eq.~\ref{3pt} can be analytically evaluated in
the slow-roll limit. Ignoring $\epsilon^3$ terms in Eq.~\ref{3pt}, gives us the
following result that was first derived by Maldacena\cite{maldacena}.
\begin{eqnarray}
\left\langle\zeta_{\bold{k}_1}\zeta_{\bold{k}_2}\zeta_{\bold{k}_3}\right\rangle
= (2\pi)^7\delta^3(\bold{k}_1+\bold{k}_2+\bold{k}_3)\frac{(P_k^{\zeta})^2}{\prod_ik_i^3} \mathcal{A}\label{3A} \\
\mathcal{A} =\frac{\eta^*-\epsilon^*}{8}\sum_i k_i^3 +
\frac{\epsilon^*}{8}\sum_{i \ne j} k_ik_j^2 + \frac{\epsilon^*}{K}\sum_{i>j}
k_i^2k_j^2\quad \label{A3p}
\end{eqnarray}
where $*$ denotes the $\epsilon$ and
$\eta$ values at the horizon crossing. The quantity $\mathcal{A}$ is a convenient
measure of non-Gaussianity in the perturbation field. The relationship between
$\mathcal{A}$ and the bispectrum is given by
\begin{equation}
B_{\zeta}(k_1,k_2,k_3)=(2\pi)^4\frac{(P_k^{\zeta})^2}{\prod_ik_i^3}
\mathcal{A}.
\end{equation}

The bispectrum and $\mathcal{A}$ are general measures of non-Gaussianity;
however, both these quantities are highly scale dependent. Thus, over the recent years 
$f_\textsc{NL}$, a local dimensionless nonlinearity parameter that is also independent 
of scale, has become a widely used measure of non-Gaussianity\cite{Komatsu1}.
One can define a generalized $f_\textsc{NL}$ for the general kind of 
non-Gaussianity by the following equation, that also has
the advantage of being nearly scale independent.
\begin{equation}
f_\textsc{NL}\equiv-\frac{10B_{\zeta}(k_1,k_2,k_3)\prod_i
k_i^3}{3(2\pi)^4(P_k^{\zeta})^2\left(\sum_i  k_i^3\right)}
\label{fg}
\end{equation}
The measurement of $f_{NL}$ and bispectra for particular
triangle configurations in the data has been used to
restrict specific (local, equilateral and orthogonal)
primordial bispectrum ampltitudes, for instance for Planck data in \cite{Planck2013ng,Planck2015ng}. At the same time these 
efforts have not to date detected primordial non-Gaussian
features in CMB data.

In this paper we focus on real space
measures of non-Gaussianity such as high order moments of the field and its derivatives and geometrical descriptors expressed as expansions in
these moments. For the first perturbative level of deviation from the Gaussian limit, only
cubic moments are involved and they are given by k-space integrals
of the bispectrum (see \cite{Pogosyan2} for exact relations). Thus, the
real space approach provides restrictions on different combination of bispectrum amplitudes, however what is more important, it links primordial
non-Gaussianity to observable quantities where the estimation from data has completely different systematics and noise properties.
Using different observables and cross validating the results
will be critical to detect weak signal in a convincing way,
especially of such a
complex effect as non-Gaussianity, which has a very wide space of
parameters and signatures.


\section{Numerical Technique} 
\subsection{Calculation of Mode Function and Power Spectrum} 
To calculate the power spectrum of scalar perturbations we need
to solve the background equations of motion Eqs.~\ref{bg} and the Mukhanov
equation Eq.~\ref{Mukha} that can also be written in the following form
\begin{eqnarray}
\dot{H} &=& -\frac{\dot{\phi}^2}{2} \notag  \\
\ddot{\phi}&=&-3H\dot{\phi}-V_{,\phi} \notag  \\ 
v''_k&=&-\left(k^2-\frac{z''}{z} \right)v_k\text{, } z = a\sqrt{2\epsilon}
\end{eqnarray}
where primes $'$ donote derivatives with respect to conformal time. These are
coupled differential equations with the first two representing the background and the
last equation for scalar perturbations.  Numerically, it is more convenient to
work out the differential equation for $u_k$ rather than $v_k$ since we finally
require $u_k=v_k/z$ to calculate power spectrum and bispectrum. We can shift for conformal time $\tau$ to the number of e-folds $n=\ln(a)$ variable and write the above equations in numerically more efficient form. Thus,
we convert the Mukhanov equation to the perturbation equation for $u_k(n)$ as a function of $n$.
Similarly we can write the above equations of motion as function of $n$ as below,
\begin{eqnarray}
H_{,n}&=&-H\frac{\phi_{,n}^2}{2}, \\
\phi_{,nn}&=&-(3-\epsilon)\phi_{,n}-V_{,\phi}/H^2,\\
{u_k}_{,nn}&=& -(3-\epsilon+\eta){u_k}_{,n} -\frac{k^2}{(aH)^2} u_k .
\end{eqnarray}
where subscripts `$_{,n}$' denote derivatives with respect to $n=\ln(a)$.
We solve these equations numerically starting mode evolution
deep inside the horizon and choosing the Bunch-Davies vacuum for the initial
conditions.

The above equations of motion are for single-field inflation with standard
kinetic term with any potential $V(\phi)$. In this paper we 
specifically studied quadratic inflation $V(\phi)=\frac{1}{2}m^2\phi^2$ as our
base model with mass of the inflaton field $m=6.125\times10^{-6}$ that gives us
the correct value of $A_s=2.215\times 10^{-9}$ at the pivot scale
$k=0.05~Mpc^{-1}$ \cite{Planck15}. We then studied a variation of the base model
that has step like feature added to the quadratic potential, as an example of
models with localized sharp potential changes.
The potential for this model is given by
$V_{step}(\phi)=\frac{1}{2}m^2\phi^2\left ( 1+ c\tanh\frac{\phi-\phi_s}{d}\right)$ with the same mass parameter $m$ whereas $c$ and $d$
are the height and the width of the step jump at
location $\phi_s$ \cite{XChen1,XChen3}.

\begin{figure}[htbp]
\begin{center}
\includegraphics[height=6.3cm]{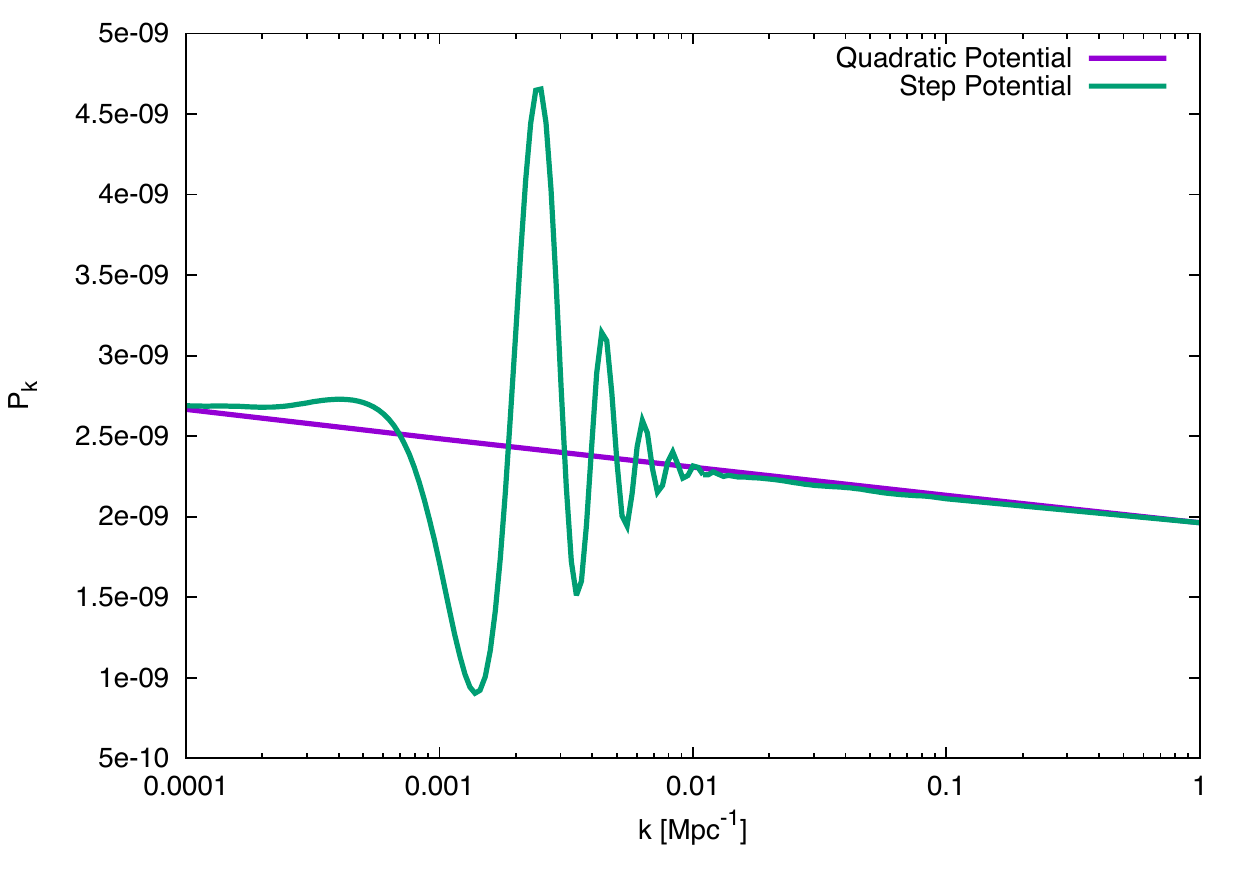}
\caption{Power spectrum of scalar perturbations $P_{\zeta}(k)$ against momenta
$k$ for quadratic potential and step potential with $c=0.002$, $d=0.02\;M_{pl}$ and
$\phi_s=15.86\;M_{pl}$. }
\label{Pk}
\end{center}
\end{figure}

In Fig \ref{Pk} we have presented the calculation of dimensionless power spectrum 
according to Eq.~\ref{PowerS}. The power spectrum is mildly dependent on
$k$ for the quadratic potential with $\frac{d\ln P_{\zeta}}{d\ln k}=-2\epsilon^*-\eta^*$. 
On the other hand for the step potential, due to the breaking of the slow-roll condition
because of a sharp step in the potential, we see an oscillating power spectrum near
the step but as we move away from the step it follows the same behavior of 
the quadratic potential (Fig. \ref{Pk}).

\subsection{Three-point function Calculation} After numerically solving the
background equations and the equation for scalar perturbations, we insert these
solutions back into Eq.~\ref{3pt} to calculate the three-point function in
momentum space. The three-point correlation function is a numerically challenging
task as it involves integrations that arise from equation (\ref{3pt}). The
integrands consist of three factors of $u_k$ or $u'_k$ multiplied by the
background factors of $a$, $\epsilon$ and $\eta$. The scalar perturbation
function $u_k$ oscillates before the horizon crossing at $\tau^*$, while
after the horizon crossing it freezes out. Thus, the integration consists of two parts, before the
horizon crossing (BHC) part and after the horizon crossing (AHC) part
\begin{eqnarray}
\int_{-\infty}^{\tau_{end}} d\tau I(\tau) =
\int_{-\infty}^{\tau^*} d\tau I(\tau)+\int_{\tau^*}^{\tau_{end}} d\tau I(\tau)
\label{integ}
\end{eqnarray}
where $\tau^*$ is the horizon crossing point of
the largest $k$ mode in the three-point correlation function and $I(\tau)$ is
the integrand of the the-point function given in Eq.~\ref{3pt} that contains
background factors and product of three oscillating mode functions. The BHC and
AHC parts of integration present different numerical challenges as the first
has growing oscillations, as $\tau$ approaches negative infinity,  while for
the AHC part one has to regularize by adding a total
derivative term in the action\cite{Arroja,Horner1}. Without adding this term in the
action, the AHC part of the integral is divergent as one of the terms
$a^3\epsilon\dot{\eta}\zeta^2\dot{\zeta}$ in the initial action grows as the
scale factor \cite{Arroja,Horner1}.

The contribution to the integral that arises from before horizon crossing poses significant technical
challenges. In conformal time the initial big bang singularity is pushed back in
conformal time to $\tau \rightarrow -\infty$. Thus, the scalar perturbations
start deep inside the horizon and keep oscillating till horizon crossing point
$\tau^*$ of the largest $k$ mode in the three-point function. Now, there are
different methods to numerically evaluate an oscillating integral over an
infinite range. If we cutoff this infinite integral to some finite value, due
to large oscillations this induces an spurious contribution of $O(1)$.
Numerically it was shown that these kind of integrals can be evaluated by
introducing an arbitrary damping factor into the integrand but this damping
factor needs to be chosen carefully\cite{XChen1}. Other techniques, such as
boundary regularization, for evaluating such integrals are even more complex
\cite{XChen3,Horner1}.

We have developed a different numerical technique, which is numerically more
robust and elegant, using the Cesaro resummation of improper series. For
oscillating integrand $I(\tau)$, the following expression gives the definition
of Cesaro integration
\begin{eqnarray}
\int_{-\infty}^{\tau^*} d\tau I(\tau)
\equiv \lim_{\tau \rightarrow -\infty} \frac{1}{\tau^*-\tau}
\int_{\tau^*}^{\tau} d\tau' \left( \int_{\tau^*}^{\tau'} d\tau'' I(\tau'')
\right)~.
\label{cesaro}
\end{eqnarray}
This gives a specific definition to the improper integral on the left-hand side, whereas the right-hand side is an average over the partial integrals that give a convergent result for a wide range of improper integrals\cite{cesa}. However, we extended this method further and we defined a higher order Cesaro integral, with double average over partial sums, to further improve the convergence defined as 
\begin{eqnarray}
\lim_{\tau \rightarrow
-\infty} \frac{1}{\tau_0-\tau} \int_{\tau_0}^{\tau} d\tau'
\frac{1}{\tau_0-\tau'} \int_{\tau_0}^{\tau'} d\tau'' \left(
\int_{\tau_0}^{\tau''} d\tau''' I(\tau''') \right)~.
\label{cesa2}
\end{eqnarray}
In our numerical program we have used this extended version of the Cesaro sum to compute BHC part.
\begin{figure}[h]
\begin{center}
\includegraphics[height=6.3cm]{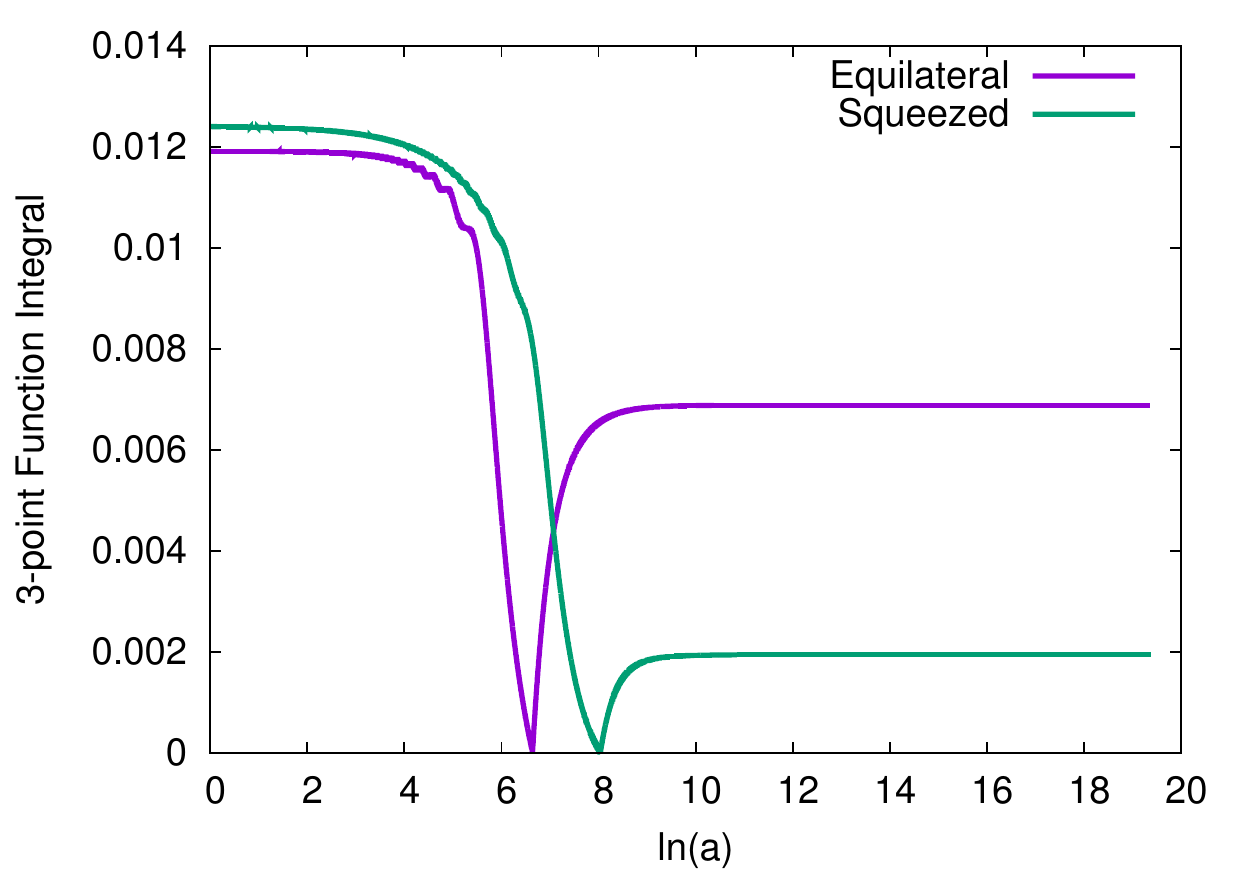}
\caption{The AHC and BHC parts of three-point function, or generalized
$f_\textsc{NL}$, are plotted against the number of e-folds $N=\ln(a)$ while integrating, for Equilateral and Squeezed triangles. The horizon crossing value $\tau^*$ correspond to the two kinks at $6.5$ and $8$ e-folds for the two triangles. The integral in Eq. \ref{3pt} is divided into two parts as given in Eq. \ref{integ}. The BHC part, to the left
of the kink, is evaluated as the integral from $\tau_*$ to $\tau \to -\infty$ ($\ln(a)\to 0$) using Cesaro resummation, It  plateaus quickly showing stability of integration technique.  The AHC part
is integrated from $\tau_*$ in the direction of increasing $\tau$, the integration result is shown to the right of the kink, and also quickly
plateaus. The generalized $f_\textsc{NL}$ is the sum of the two integration asymptote(plateau) values, at the left and the right of the plot.
}\label{uuu}
\end{center}
\end{figure}

This method quickly gives convergent results without introducing any artificial
damping factors. This can be seen in Fig.~\ref{uuu} which shows the three-point
function integral AHC and BHC results plotted against the number of e-folds (while integrating) for equilateral triangle and squeezed triangle cases. In this figure horizon
crossings occur $\tau^*$ that correspond to e-folds values of $6.5$ and $8$
for equilateral and squeezed triangle. This Fig~\ref{uuu} describe two
different integration regimes BHC $\tau<\tau^*$ and AHC $\tau>\tau^*$. In BHC
regime, we integrate in the backward direction from the horizon crossing point $\tau^*$ using the extended Cesaro Integral\cite{Junaid}. Our technique converges very
quickly as can be seen that the integral plateaus as we go 5-6 e-folds left of the
horizon crossing points. In the AHC regime $\tau > \tau^*$, we integrate in the forward
direction starting at $\tau^*$ that also plateaus soon after horizon crossing. Thus, the three-point
function integral, or generalized $f_\textsc{NL}$, is just the sum of the two
asymptote(plateau) values in the before and after horizon crossing regimes for
each kind of triangle (see Fig. \ref{uuu}).

\begin{figure}[htbp]
\begin{center}
\includegraphics[height=6.3cm]{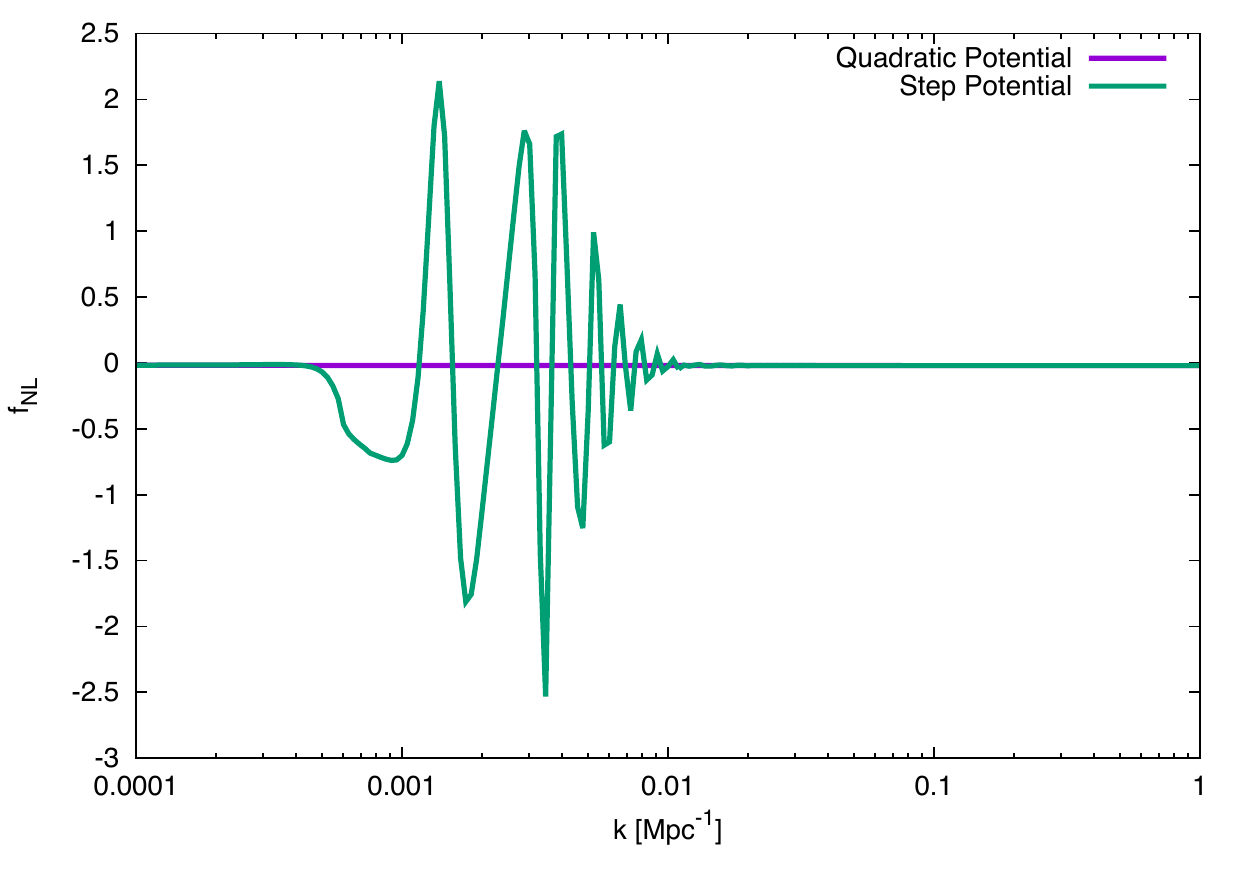}
\caption{Generalized $f_\textsc{NL}$ for the step potential model plotted against $k=k_1=k_2=k_3$ for 
equilateral triangle case with step height and width given by $c=0.002$ and
$d=0.02M_{pl}$ respectively while step is located at $\phi_s=15.86\;M_{pl}$.}
\label{fnls}
\end{center}
\end{figure}

To test our procedure, we have calculated the three-point function numerically
for the $\frac{1}{2}m^2\phi^2$ potential and compared it with the corresponding
analytical results given by Eq.~\ref{A3p}. The results show that our numerics match the analytical results with the error bar below $1\%$ for slow-roll models of inflation\cite{Junaid}. The generalised $f_{NL}$ is plotted in Fig. \ref{fnls} for the step potential that shows that our technique works even for such potentials.


\section{2D Moments for CMB maps}\label{MG}
Non-Gaussianity can also be studied
through the higher order moments of the perturbation field in configuration
space. Analysis of these moments provides a robust measure of non-Gaussianity
and has also become an important field of investigation\cite{wmap2,Park}. In
this paper we will extend our previous work on 3D calculation of moments \cite{Junaid}. 
These 2D moments will provide important information on the geometrical properties
of the CMB temperature fluctuations map, and give us 2D non-Gaussianity observables such as
extrema counts, genus and skeleton\cite{Pogosyan1}. 

To calculate the third-order moments we have to take
the inverse Fourier transform of three-point function in
momentum space. These moments are always scaled 
by the corresponding variances $\sigma$ and $\sigma_1$ 
in all physical observables such as Minkowski functionals and 
Euler characteristics.

The tiny temperature fluctuations in the CMB mark the imprints of inflation in
the early universe. The adiabatic scalar mode of these fluctuations are generated by
the quantum fluctuations of initial perturbation field $\zeta$. Thus, we will study 
how these quantum perturbations get imposed on the CMB maps.
In two-dimensional map of the sky, the temperature fluctuations 
can be related to $\zeta$ via the k-mode integral
\begin{eqnarray}
\frac{\Delta T}{T}(\theta,\varphi) &\approx&
\int \frac{d^3k}{(2\pi)^3} \zeta_\mathbf{k} \Theta(k,\mathbf{\hat{k}} \cdot \mathbf{\hat n},,\tau_{rec},\tau_{now})
\label{eq:thetadef}
\end{eqnarray}
where the transfer function $\Theta(k,\mathbf{\hat{k}} \cdot \mathbf{\hat n},\tau_{rec},\tau_{now})$ describes how the
observed temperature of the photons coming from the direction $\mathbf{\hat n}=(\theta,\phi)$ was formed from earlier
times to the moment of recombination $\tau_{rec}$ and subsequently modified to the present moment of observation
$\tau_{now}$. Full treatment requires reconstruction of the $\Theta$ transfer function from the solution of the
Boltzmann equation, given, for example by CAMB software  \cite{Lewis12,Lewis1999bs}. Here we adopt an approximate
approach, that of an instant recombination, suitable for sufficiently large scales exceeding the sound
horizon at recombination, $k \tau_{rec} c_s < 1$ ($c_s$ being the speed of sound)
\begin{eqnarray}
\Theta
& \approx &T(\mathbf{k}) e^{i \mathbf{k} \cdot \mathbf{\hat n} \tau_0}\label{premonosub}\\
T(\mathbf{k})&=&
\left[ \frac{1}{4} \Delta_\gamma + \Phi + i \; \mathbf{k} \cdot \mathbf{\hat n} \;  v_e \right]_{\tau_{rec}}
e^{-\sigma(\tau_{rec})} \nonumber \\
&+& \int_{\tau_{rec}}^{\tau_{now}} \mathrm{d} \tau (\Phi^\prime+\Psi^\prime)
e^{i \mathbf{k} \cdot \mathbf{\hat n} (\tau_{rec}-\tau)} 
e^{-\sigma(\tau)}
\label{eq:tkdef}
\end{eqnarray}
where the Newtonian potentials $\Phi$ and $\Psi$, the photon energy density fluctuations $\Delta_\gamma$ and electron
velocity potential $v_e$ 
are easily obtained from cosmological perturbation codes as in CAMB code \cite{Lewis12,Lewis1999bs}.
The above transfer function describes photons that propagated nearly freely
from the moment of their last scattering $\tau_{rec}$ over the radial
distance $\tau_0=\tau_{now}-\tau_{rec}$ 
with temperature set by the local temperature of the plasma shifted by the gravitational and Doppler shifts
at the position of the photon release and modified at later time due to propagation in time variable potential (
often called integrated Sachs Wolfe (ISW) effect).
Since the transfer function is defined after the primordial $\zeta_k$ amplitude is factorized out, the appropriate
normalization of perturbation potentials on superhorizon scales $k \to 0 $ is $ \frac{1}{4} \Delta_\gamma - \Psi \to 1$.
This corresponds to $T(k \to 0 )=-1/5$ in Einstein-de Sitter universe \cite{baumann2,dodelson,ALwis}.
In Eq.~(\ref{eq:tkdef}) $\sigma(\tau)$ is the optical depth due to late-time scattering along the photon path from
the moment $\tau$ to the observer at $\tau_{now}$. This optical depth predominantly accumulates from
the time reionization of the Universe at $\tau_{reion} $ to $\tau_{now}$ 
and according to the latest Planck Collaboration analysis amounts to $\sigma(\tau_{rec})=0.066$ \cite{Planck15params}.


Note that the integral over full
sky of temperature fluctuations is zero by definition, so the monopole term is unobservable and must be subtracted out
prior to the calculation of any configuration space statistics  \cite{Eriksenetal2004,ZHou2,ZHou,Larson}. Given that
the angular dependent part of the ISW integral and Doppler term contribute little to the monopole, the monopole
subtraction can be achieved by simply replacing $
e^{i \mathbf{k} \cdot \mathbf{\hat n} \tau_0} \to 
e^{i \mathbf{k} \cdot \mathbf{\hat n} \tau_0} - j_0 (k \tau_0)$ in Eq.~(\ref{premonosub}).

Let us look how one can calculate the moments of $\widetilde{\zeta}(\bold{\hat{n}}) =\frac{\Delta T(\bold{\hat{n}})}{T}$ on the 2D sky. After subtracting
unobservable monopole contribution 
\begin{eqnarray}
\widetilde{\zeta}(\bold{\hat{n}}) &=& \int \frac{d^3k}{(2\pi)^3} \zeta_k  T(k) W(kR)\left[ e^{i \mathbf{k} \cdot \mathbf{\hat n} \tau_0} - j_0\left( k \tau_0 \right) \right]~~
\label{monosub}
\end{eqnarray}
Here we have introduced the very important quantity for the analysis of the data, the window function $W(k R)$.
Configuration space statistics are always measured in the experiment when the data field is suitably smoothed.
Dependence of the statistics on the smoothing scale is an important informative ingredient for matching
observational data to theoretical predictions. In our case the smoothing of the temperature field is done on the sky,
and $W(k R)$ is the Fourier space response of this angular (or multipole) smoothing. In general smoothing
may contain both low-pass component, that suppresses small scale contributions, and a high-pass part that suppresses
very long angular variations. While the latter can be useful to eliminate contribution of low multipoles,
in particular the dipole part of the temperature map, in this paper we limit our analysis to low-pass filtering.
We shall consider the window function to be Gaussian with cutoff scale $R$, $W(k R) = e^{-k^2 R^2/2}$.
Note that in general, as a rough rule of thumb, one can use the correspondence $R = \tau_0/\ell$
between the real space scale $R$ and the angular multipole smoothing scale $\ell$ on a sphere of radius $\tau_0$.

The variances of the temperature fluctuations and it gradients are then given by
\begin{eqnarray}
\sigma^2 &=& \left< \widetilde{\zeta}^2 \right> = \int d\ln(k)
\label{sigmatilde} \\
&& \quad \times P_\zeta(k) T^2(k) W^2(kR)\left( 1 - j^2_0(k\tau_0)\right)
\notag \\
\sigma_1^2 &=& \left< (\nabla \widetilde{\zeta})^2 \right> = \int k^2 d\ln(k)
\label{sigma1tilde} \\
&& \quad \times P_\zeta(k) T^2(k) W^2(kR)\left( 1 - j^2_0(k\tau_0)\right)
\notag 
\end{eqnarray}
which we denote for brevity as, respectively, $\sigma^2$ and $\sigma_1^2$.

Next we calculate the cubic moments with monopole subtraction. A general solution for calculation of these moments is given in Appendix A. 
We define for brevity the double integral over wave vector magnitudes as
\begin{eqnarray}
\int d\Omega_{ps}&=&\int \frac{k_1^2dk_1}{4\pi^2}\frac{k_2^2dk_2}{2\pi^2}d\cos\theta B_\zeta(k_1,k_2,|\bold{k}_1+\bold{k}_2|)\notag\\
&~&\quad \times \prod_{i=1}^3 W(k_i R) T(k_i),~\bold{k}_3=-\bold{k}_1-\bold{k}_2.
\end{eqnarray}
Using this notation we obtain the following expression for the cubic
moment \cite{PPG11,Junaid}
\begin{eqnarray}
\left< \widetilde{\zeta}^3 \right> &=& \int d\Omega_{ps}\Bigg(1 +2j_0(k_1\tau_0)j_0(k_2\tau_0)j_0(|\bold{k}_3|\tau_0)\notag\\ &~&\quad - \frac{3}{2} \left( j^2_0(k_1\tau_0) +j^2_0(k_2\tau_0)  \right)  \Bigg), \label{tilde_x3}
\end{eqnarray}
Similarly, we found that
\begin{eqnarray}
&\left< \widetilde{\zeta}^2 \Delta\widetilde{\zeta}\right>&
=-\int d\Omega_{ps} \frac{2}{3}k_2^2 \tau_0^2\Bigg( 1 -2\left(j_0(k_1\tau_0) \right. \notag
\\&~& \left.+j_2(k_1\tau_0)P_2(\cos\theta) \right)+ j_0(k_1\tau_0)\label{tilde_x2I} \\
&~&\times j_0(|\bold{k}_3|\tau_0)\left( j_0(k_2\tau_0)+j_2(k_2\tau_0)\right)\Bigg)\notag.
\end{eqnarray}
and 
\begin{eqnarray}
\left< (\nabla \widetilde{\zeta})^2 \Delta\widetilde{\zeta}\right>
&=&  \int d\Omega_{ps} \frac{4}{3} \left(  \frac{k_{1}^4+k_2^4}{10}\right.\label{tilde_q2I}\\&~&\qquad   - \left. \frac{k_{1}^2 k_{2}^2}{3}\left(1+ \frac{P_2(\cos\theta)}{5}\right) \right) \tau_0^4.\notag
\end{eqnarray}

Using the above expressions for two-dimensional moments we have calculated these
moments for the quadratic potential as well as for the step and $\lambda\phi^4$
potentials. 
We have used the values of $\tau_{now}=14362\;\mathrm{Mpc}$, 
$\tau_{rec}=284.95\;\mathrm{Mpc}$ and $\tau_0=14077\;\mathrm{Mpc}$
as given by CAMB software
for the best fit cosmological parameters from the Planck Collaboration results\cite{planck1,planck4}.
Our approximate treatment of temperature fluctuations is applicable
for $k \tau_{rec} c_s < 1$, i.e $k \tau_0 < \tau_0/(\tau_{rec} c_s) \approx 94 $ with $c_s \approx 0.526\; c$.
For numerical reasons, the integration over wave numbers have been limited to the range ($k_{min}=0.3/ \tau_0$, $k_{max}=300/ \tau_0$) for $k$ integrals,
which was checked not to affect our results significantly, due to already
present monopole cutoff at low $k < 1/\tau_0$ and the use of low-pass Gaussian
smoothing with sufficiently large scale R, $R \; k_{max} > 1$.

We shall quote the results for the normalized 2D cubic moments
$S_2=\left< \widetilde{\zeta}^3\right>/\sigma^4$,
$T_2=\left< \widetilde{\zeta}^2 \Delta\widetilde{\zeta}\right>/\sigma^2\sigma_1^2$
and
$U_2=\left< (\nabla \widetilde{\zeta})^2 \Delta\widetilde{\zeta}\right>/\sigma_1^4$.
Such normalized moments are scale independent in the second
order of perturbation theory for a scale-free power spectrum $P_\zeta(k)$
but are inverse proportional to the linear change of the amplitude 
of $\zeta$ field. Non-Gaussian corrections to  
geometrical configuration space statistics
such as Minkowski functionals are proportional to $\sigma S_2$, $\sigma T_2$ and
$\sigma U_2$ and do not dependent on the linear field amplitude.

In Table~\ref{MonoMoments} the magnitudes of
normalized moments and average $f_{NL}$ value for different inflationary models are compared.
For the table we have chosen the smoothing length to be 
$R=422\; Mpc = 0.03\;\tau_0$. This real-space scale roughly corresponds to
$\ell=33$ multipole. The variance at this smoothing is 
$\sigma \approx 3 \times 10^{-5}$.

\begin{table}[htb]
\begin{center}
\begin{tabular}{ccccc}
\hline\hline
& $\hspace{0.1 cm} $Inflationary Model & $\hspace{0.1 cm} S_2$ & $\hspace{0.2 cm} T_2$ & $\hspace{0.2 cm} U_2$\hspace{0.5 cm} $f_{NL}$\hspace{0.3 cm} \\ \hline
& $\hspace{0.1 cm} \frac{1}{2}m^2\phi^2$ Potential & $\hspace{0.1 cm} 0.147$ & $\hspace{0.2 cm} -0.171$ & $\hspace{0.2 cm} 0.017$\hspace{0.3 cm} $0.02$\hspace{0.3 cm} \\ \hline
& $\hspace{0.1 cm} \lambda\phi^4$ Potential & $\hspace{0.1 cm} 0.418$ & $\hspace{0.2 cm} -0.493$ & $\hspace{0.2 cm} 0.076$\hspace{0.3 cm} $0.03$\hspace{0.3 cm} \\ \hline

& $\hspace{0.1 cm} \text{Step}_{c=0.002,~d=0.02M_{pl}}$ & $\hspace{0.1 cm} 1.806$ & $\hspace{0.2 cm} -2.312$ & $\hspace{0.1 cm} 0.894$\hspace{0.3 cm} $2.0$\hspace{0.3 cm} \\  \hline
& $\hspace{0.1 cm} \text{Step}_{c=0.01,~d=0.01M_{pl}}$ & $\hspace{0.1 cm} 4.383$ & $\hspace{0.2 cm} -5.146$ & $\hspace{0.0 cm} 6.02$\hspace{0.3 cm} $50$\hspace{0.3 cm}\\ \hline
& $\hspace{0.1 cm} \text{Step}_{c=0.1,~d=0.01M_{pl}}$ & $\hspace{0.1 cm} -35.1$ & $\hspace{0.2 cm} 141.5$ & $\hspace{0.1 cm} 156.7$\hspace{0.2 cm} $1200$\hspace{0.3 cm}\\ \hline \hline
\end{tabular}
\caption[Moments in 2D for different Inflationary models]{The moments of CMB temperature fluctuation after the monopole subtraction for the smoothed field
with $R=\tau_0/30$ with $5\%$ numerical uncertainty.  The value $\tau_0=14362 Mpc$ has been used and for the models with the step potential the step is located at $\phi_s=15.86\;M_{pl}$. The approximate peak value of generalized $f_{NL}$ computed from Eq.~(\ref{fg}) for equilateral configuration $k_1=k_2=k_3$ is shown in the last column for each model for comparison. }
\label{MonoMoments}
\end{center}
\end{table}

We see that the magnitude of cubic moments is small for the classical models
with smooth potential, with $\lambda \phi^4$ potential generating somewhat
larger non-Gaussianity than the quadratic one.
Inflationary models with the step potential have the magnitude
of the moments $10$ to $1000$ times larger which enhances the possibility of non-Gaussian features to be observable. {For non-Gaussianity sourced by the second-order perturbations, the normalized moments reflect, first of all, the structure of the underlying model,
as well as, if the model is not scale invariant, the smoothing procedure that selects the contributing range of wavenumbers. To get a feel for the meaning of the magnitudes of these moments, let us refer to
another well known perturbative theory, that of gravitational instability in the later matter dominated stage of Universe evolution that leads to structures in the universe that we observe.
At the second order of perturbation theory gravitational instability generates $S_3=34/7 \approx 4.86$ for the field of density perturbations $\delta=\Delta\rho/\bar\rho$
(if we neglect effects of scale dependence of the variance, see, e.g., \cite{Bernardeau1994}), which is the value similar to our $c=d=0.01$ step case. 
The difference is that as density perturbations grow, the variance of $\delta$ increases and the non-Gaussian features $\sim \sigma S_3$  become very pronounced when mildly non-linear regime $\sigma > 0.01$ is reached.
In the CMB problem, $\sigma$ of temperature fluctuations remains small, at $3 \times 10^{-5}$ level suppressing the expected non-Gaussian corrections.
}

\begin{figure}[tbp]
\begin{center}
\includegraphics[height=6.3cm]{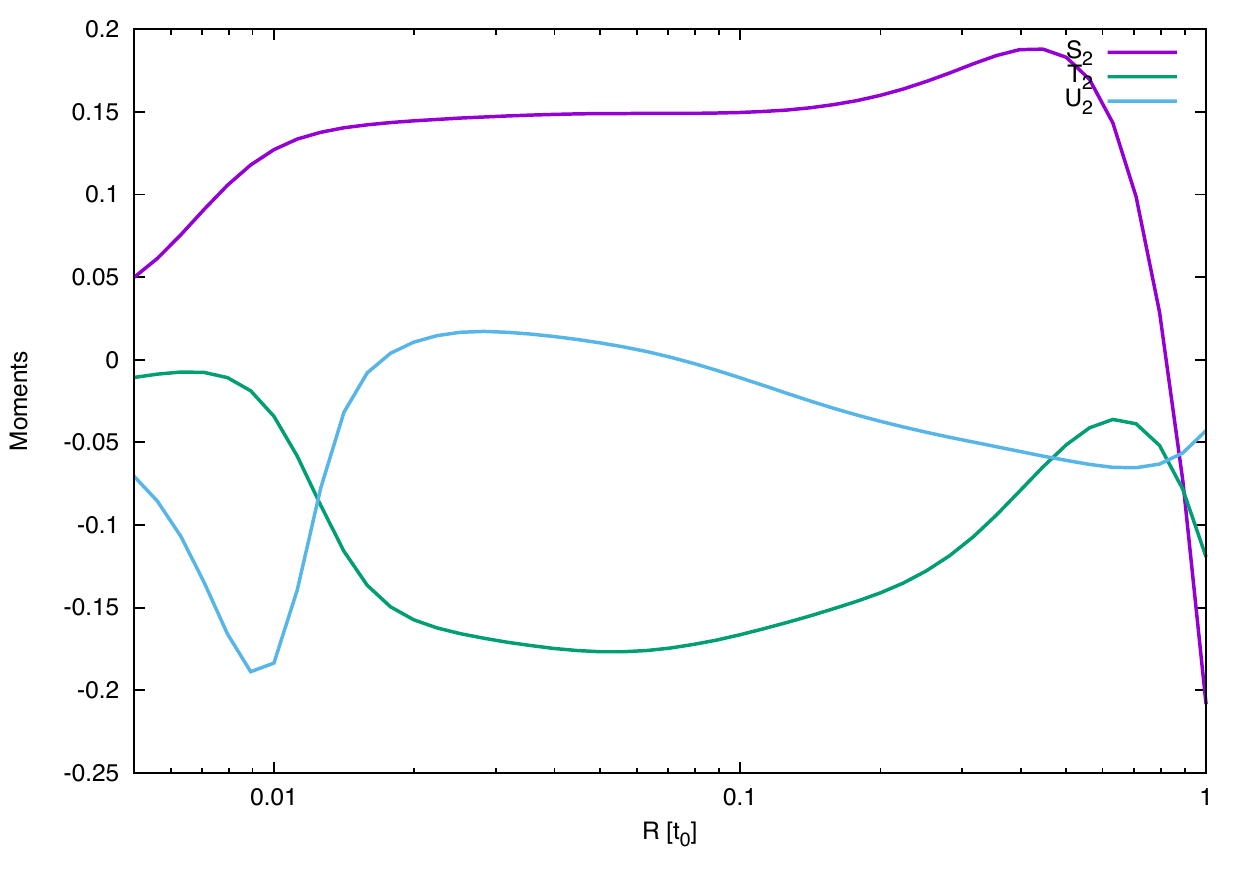}
\caption{Moments $S_2$, $T_2$ and $U_2$ plotted as a function of smoothing scale $R$ for the quadratic potential model of inflation with transfer function generated by CAMB code.}
\label{Mq}
\end{center}
\end{figure}

Now let us study 2D moments as functions of smoothing cutoff $R$.
In Fig.~\ref{Mq} we plot them for quadratic potential.
In this model $\zeta$ perturbations from inflation are nearly scale-free 
and scale dependent features of non-Gaussian cubic moments come exclusively from
the response of CMB temperature fluctuations.
As can be seen in this
figure, at sufficiently large $R$ (near the horizon scale) the moments
reflect the integrated Sachs-Wolfe effect as well as are affected by the
monopole subtraction. At scales $0.02 < R/\tau_0 < 0.2$ where higher 
spatial harmonics dominate, but the transfer function $T(k)$ is roughly flat,
$S_2$ and $T_2$ tend to a plateau values equal to the values of 3D
$\zeta$-field studied in Paper I\cite{Junaid}. To be exact, since normalized moments 
inversely change with the amplitude of the field, and $T(k) \ne 1$, the
individual values of $S_2$ and $T_2$ differ from $S_3$ and $T_3$, but their
ratio is preserved, $S_2/T_2 \approx S_3/T_3 \approx -0.8 $.
At the same time the $U_2$ moment that includes the second derivatives of the field,
is more sensitive to power increase at small scales and
continues changing through this scale range.
At $R < 0.02 \tau_0$ we start to observe the effect of the first peak in
the CMB transfer function, which increases $U_2$, but decreases $S_2$ and $T_2$
moments.

\begin{figure}[htbp]
\begin{center}
\includegraphics[height=6.3cm]{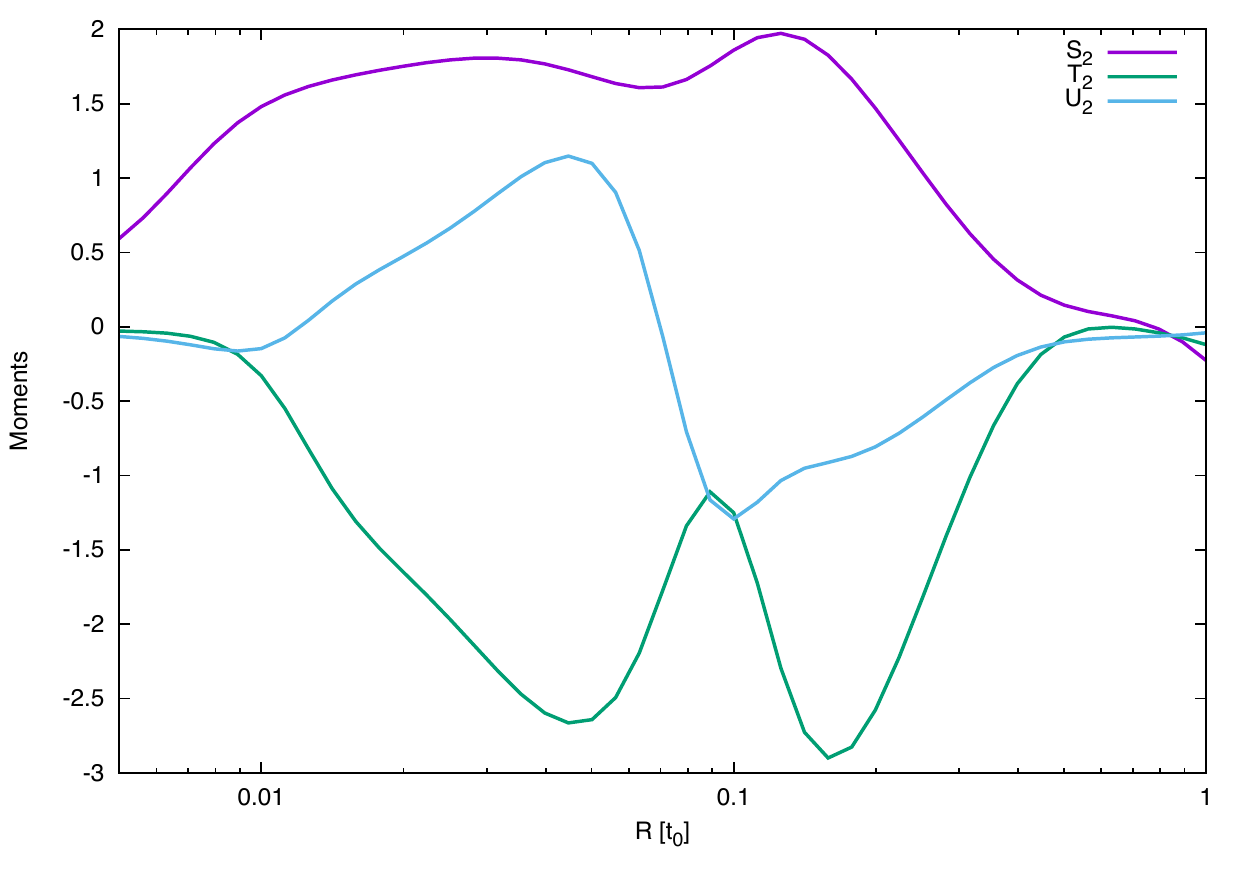}
\caption{Moments $S_2$, $T_2$ and $U_2$ plotted as a function of smoothing scale $R$ for the step model with $c=0.002$, $d=0.02M_{pl}$ and $\phi_s=15.86\;M_{pl}$ that corresponds to scale $R_s\approx\tau_0/2$.}
\label{Ms1}
\end{center}
\end{figure}

For the step potential $R$ dependence is much more informative.
In Fig.~\ref{Ms1} 
correspondent moments are plotted for the model with a step
in the inflaton potential positioned at $\phi_s$ that corresponds to the wave
vector $k_s^{-1} \equiv R_s=0.5 \tau_0$ (scale that crosses the horizon on
inflation 
exactly when the field goes through the jump in the potential, $\phi=\phi_s$).
The break in the slow-roll behavior leads to a potentially large non-Gaussian
contribution over the range of $k$ modes on the ultraviolet side of the
step scale. In these calculations,  where we used $c=0.01$ and $d=0.01M_{pl}$,
the affected k-range spans nearly two decades from $k\approx 1/R_s$ to
$k\approx 50/R_s$ and leads to complicated, often oscillatory, behavior
of the moments for smoothing scales $ R_s/50 \lesssim R \lesssim R_s$.
The exact dependence of $S_2,T_2$ and $U_2$ on $R$ in this range
is sensitive to parameters of the
model and thus gives us a discriminating test. This sensitivity will be
reflected as well in Minkowski functionals for CMB maps.

\begin{figure}[htbp]
\begin{center}
\includegraphics[height=6.3cm]{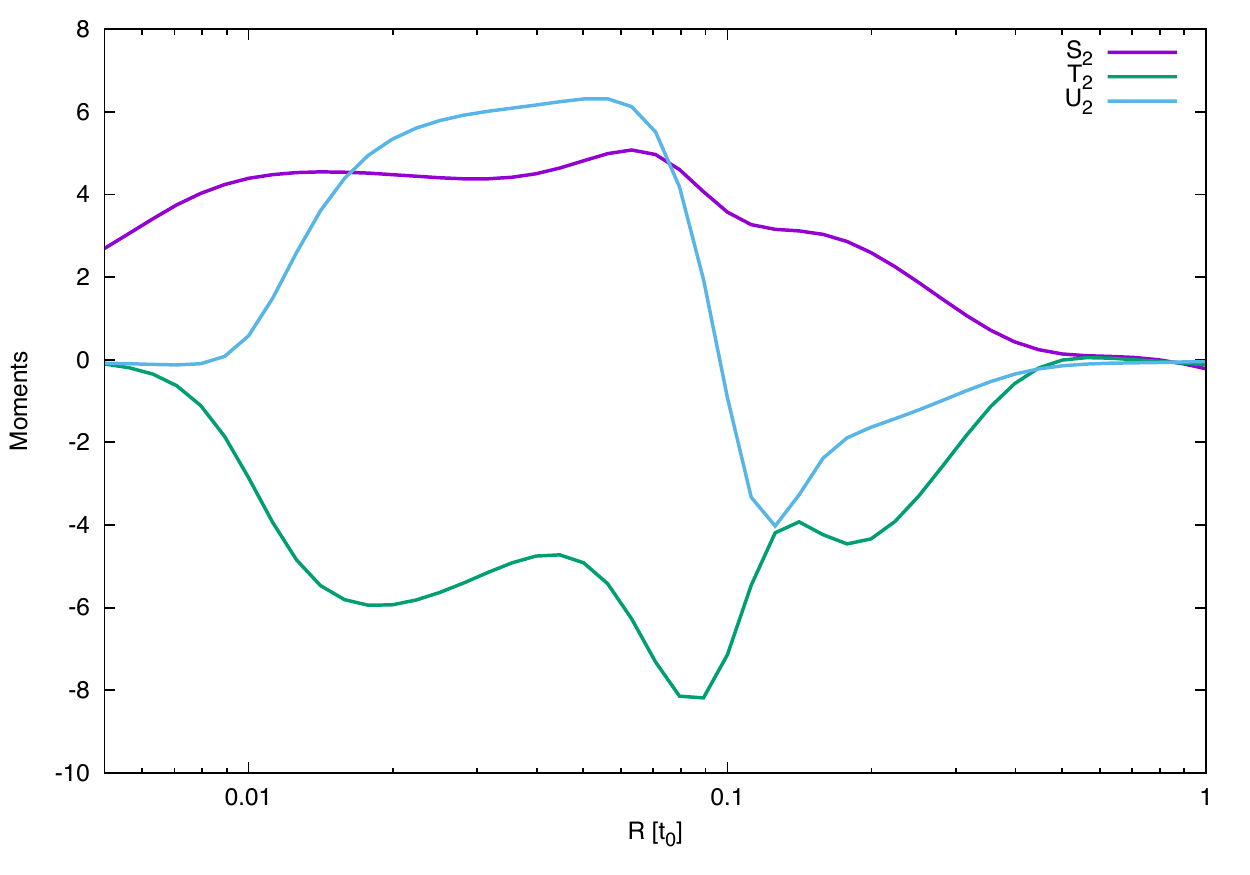}
\caption{Moments $S_2$, $T_2$ and $U_2$ plotted as a function of smoothing scale $R$ for the step model with $c=0.01$, $d=0.01M_{pl}$ and $\phi_s=15.86\;M_{pl}$ that corresponds to scale $R_s\approx\tau_0/2$.}
\label{Ms2}
\end{center}
\end{figure}

In Fig.~\ref{Ms2} we demonstrate the sensitivity of non-Gaussian features to
the parameters of the step, showing the results for $c=0.01$, $d=0.01 M_{pl}$
and the same position of the step as in Figure~\ref{Ms1}.
Both the magnitude of the cubic moments and the pattern of their $R$ dependence
is changed although the range of affected scales remained almost the same as
in Figure~\ref{Ms1}.

\begin{figure}[htbp]
\begin{center}
\includegraphics[height=6.3cm]{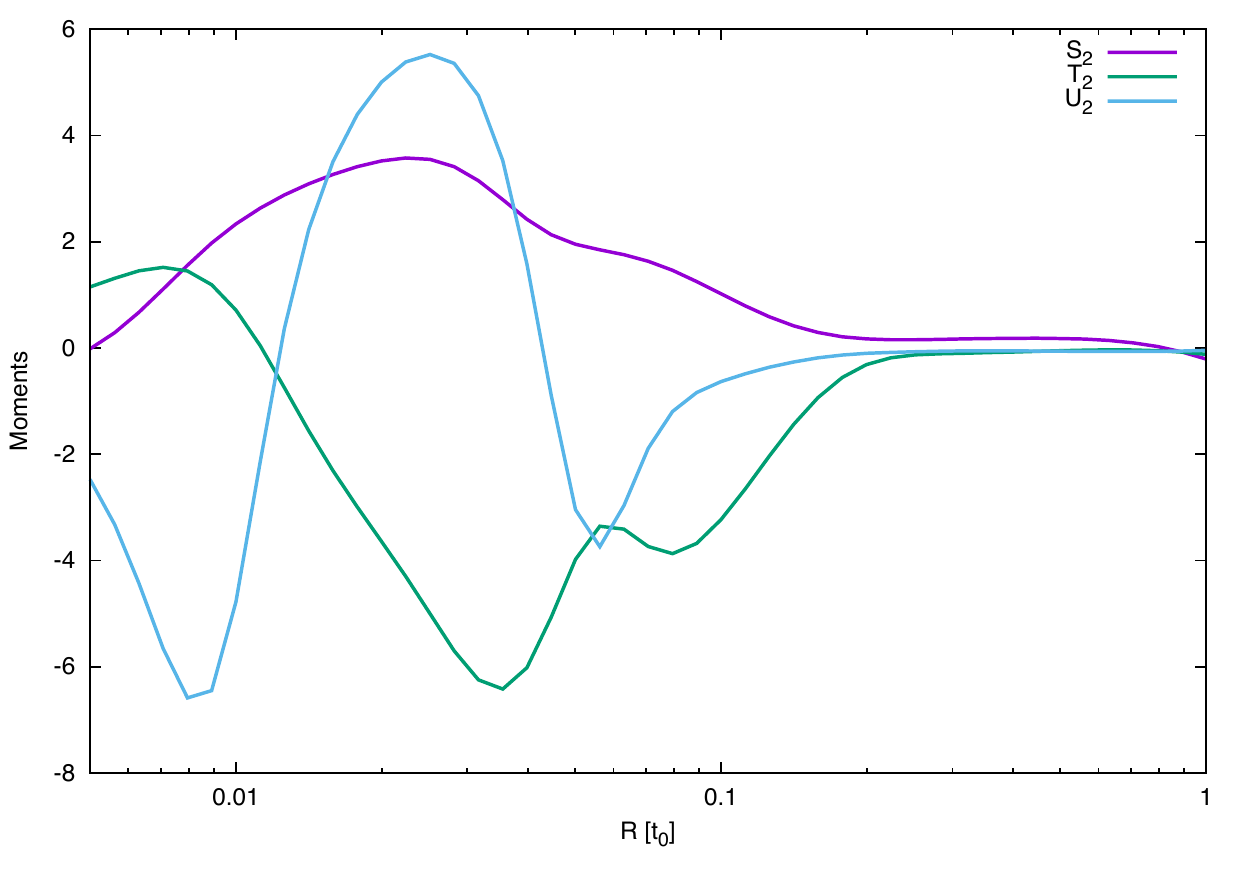}
\caption{Moments $S_2$, $T_2$ and $U_2$ plotted as a function of smoothing scale $R$ for the step model with $c=0.01$, $d=0.01M_{pl}$ and $\phi_s=15.76\;M_{pl}$ that corresponds to  scale $R_s\approx\tau_0/10$.}
\label{Ms3}
\end{center}
\end{figure}

For $R \gtrsim R_s$, we obtain the same moment values as in the base quadratic
slow-roll potential since all modes affected by the step are filtered out. 
This is illustrated in Figure~\ref{Ms3} where the position of the break
is shifted to shorter scales, $R_s\approx \tau_0/10$.
When $R \lesssim R_s/50$ we are adding $k$-modes that are less and less
affected by the step and eventually the moments values reach a plateau.

\begin{figure}[htbp]
\begin{center}
\includegraphics[height=6.3cm]{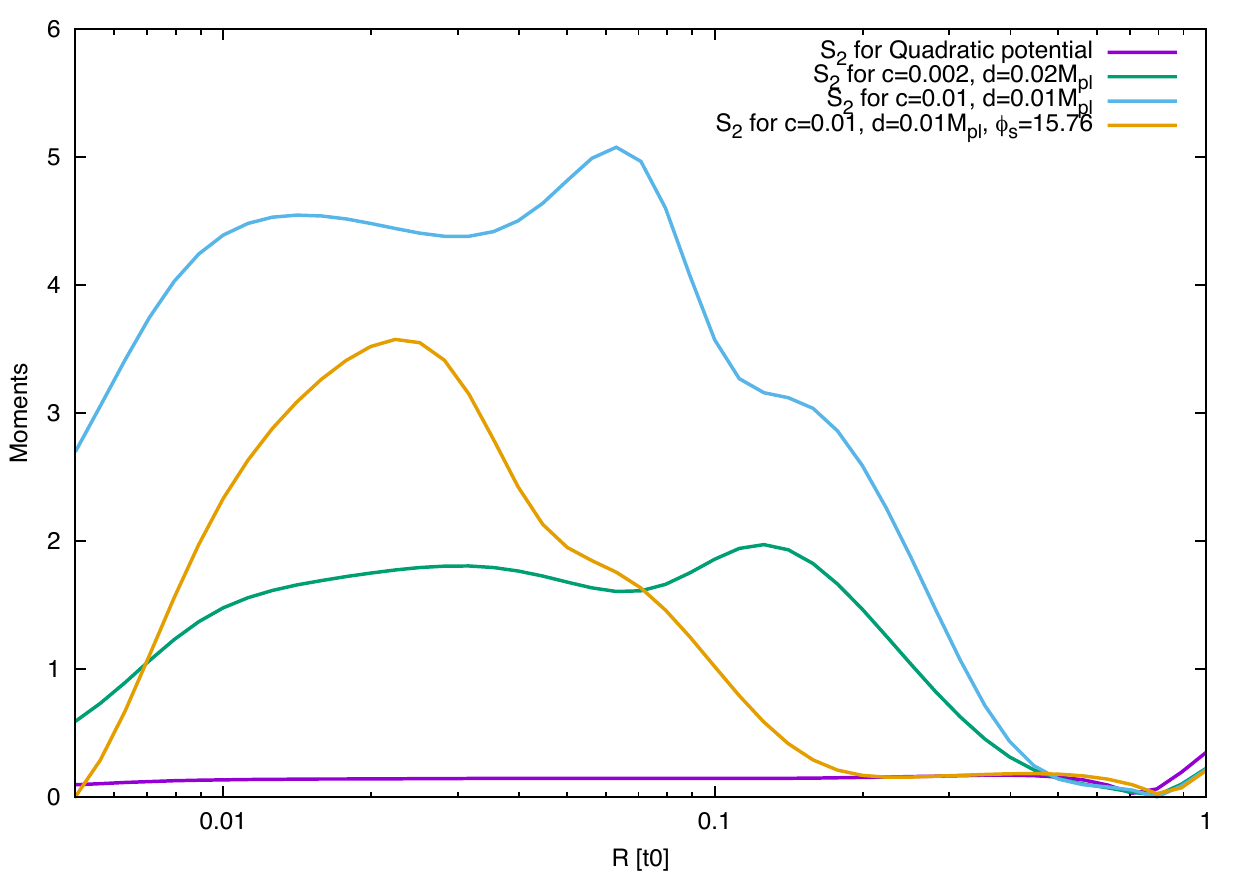}
\caption{Moments $S_2$ are plotted as a function of smoothing scale $R$ for different models and parameters for comparison.}
\label{Msc}
\end{center}
\end{figure}

To summarize, in inflationary models with step potential,
the non-Gaussian cubic moments of CMB maps show strong variation on smoothing
scale $R$. The details of $R$ behavior depend upon the location of the step
$\phi_s$ in the potential as well as the height $c$ and width $d$ parameters
of the step as depicted in Fig. \ref{Msc}. This can be used to determine the parameters of inflationary
model from the analysis of CMB maps. In particular, the position of the step,
and therefore the energy scale where the inflation potential has a feature,
is readily observed as the value of the smoothing scale below which
the region of enhanced non-Gaussianity lies. Determining $c$ and $d$ parameters
requires more detailed matching with theoretical templates.


\section{Geometrical Statistics and Minkowski Functionals} 
The Minkowski functionals up to first non-Gaussian corrections 
can be calculated for two-dimensional random fields from the 
cubic moments $S_2$, $T_2$ and $U_2$ \cite{Matsubara2,PPG11,Pogosyan2}.
Even though MFs contain formally the same information as the 
higher moments of the field and its derivative, in practice MFs
provide more robust way to recover this information from
observational and simulated data than the direct evaluation of higher
order moments. The main reason for this is that the direct moment estimation
is highly sensitive to high/low outliers in the map, and as such is very noisy
in the presence of glitches in the maps, in application to CMB for example,
residuals of point source subtraction. In contrast, MFs are functions of the threshold $\nu$, fit to which will be dominated
by the range of thresholds where the signal is the least noisy.
Reconstruction of the moments
then involves determining coefficients of decomposition of the Minkowski 
functional curves into the orthogonal basis of Hermite polynomials,
which is usually a stable procedure with better accuracy of the result.

Let us first look at Minkowski functionals for the maps smoothed with
the fixed filter. We took $R=\tau_0/50$ (i.e $\ell \approx 50$ in angular scale)
for this illustration.
The plots for non-Gaussian corrections to MFs in this section are all 
normalized as
in the Planck Collaboration paper \cite{planck3}, namely we divide them by
$\nu$-independent prefactor in the Gaussian limit of the correspondent MF. 

\begin{figure}[htbp]
\begin{center}
\includegraphics[height=5.2cm]{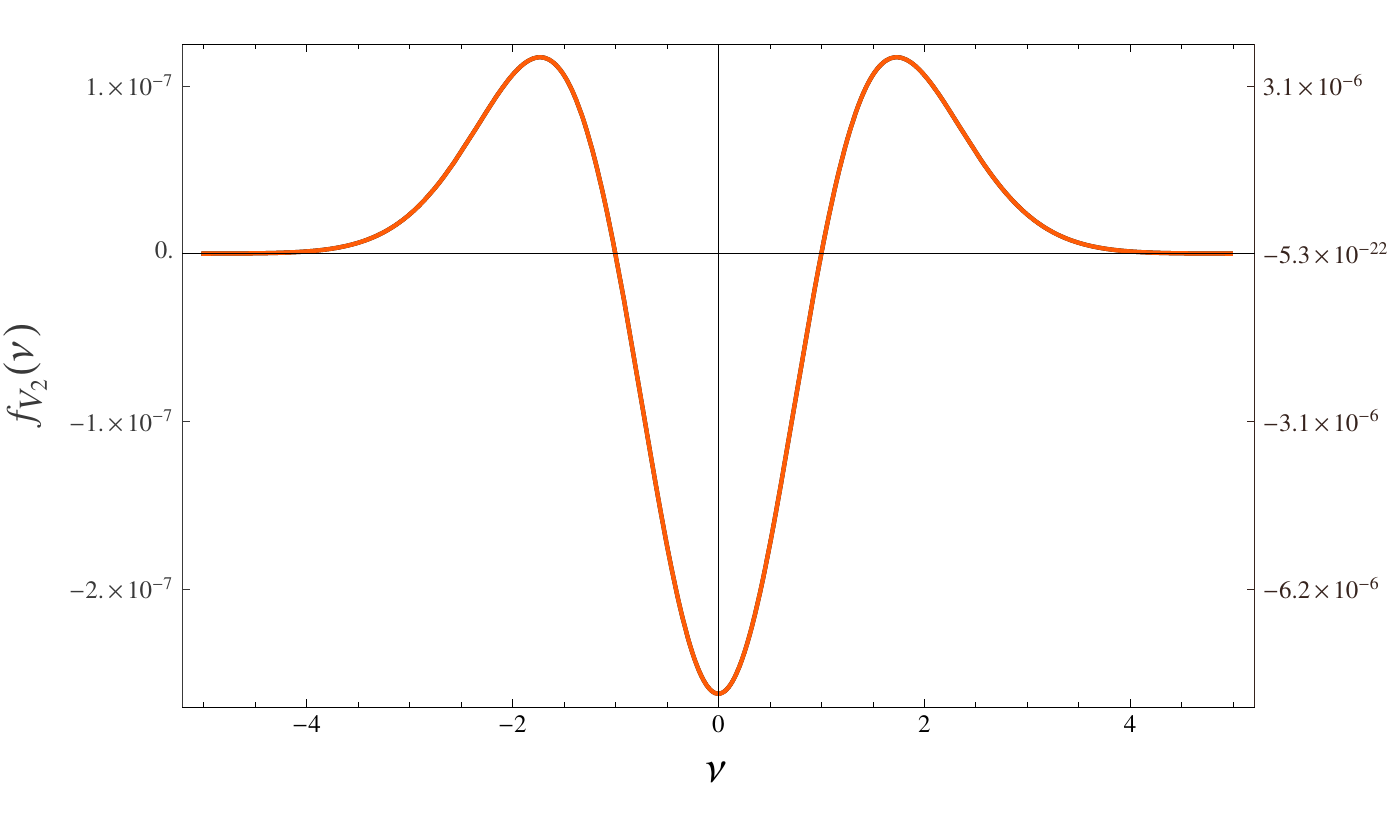}
\caption{The non-Gaussian correction to filling factor $f_{V_2}$ as a function of threshold $\nu$ for both quadratic and step potentials with $c=0.01$, $d=0.01M_{pl}$. Left axis for the quadratic potential and right for the step potential while the smoothing scale is $R=\tau_0/50$.}
\label{f2v}
\end{center}
\end{figure}

The simplest 2D Minkowski functional is the filling factor
\begin{eqnarray}
f_{V_2}(\nu)=\frac{1}{2}\mathrm{Erfc}\left( \frac{\nu}{\sqrt{2}} \right)
+\frac{e^{-\frac{\nu^2}{2}}}{\sqrt{2\pi}}\sigma H_2(\nu)\frac{S_2}{6} +O(\sigma^2).~
\label{eq:fv}
\end{eqnarray}
The first order correction in $\sigma$ to the filling factor in two-dimensions is plotted in Fig.~\ref{f2v} for the quadratic and the step potentials. 
The shape of this correction is universal but the magnitudes give 
direct access to $S_2$. In case the of the step potential it is factor of $30$
higher than for the quadratic potential.

\begin{figure}[htbp]
\begin{center}
\includegraphics[height=5.2cm]{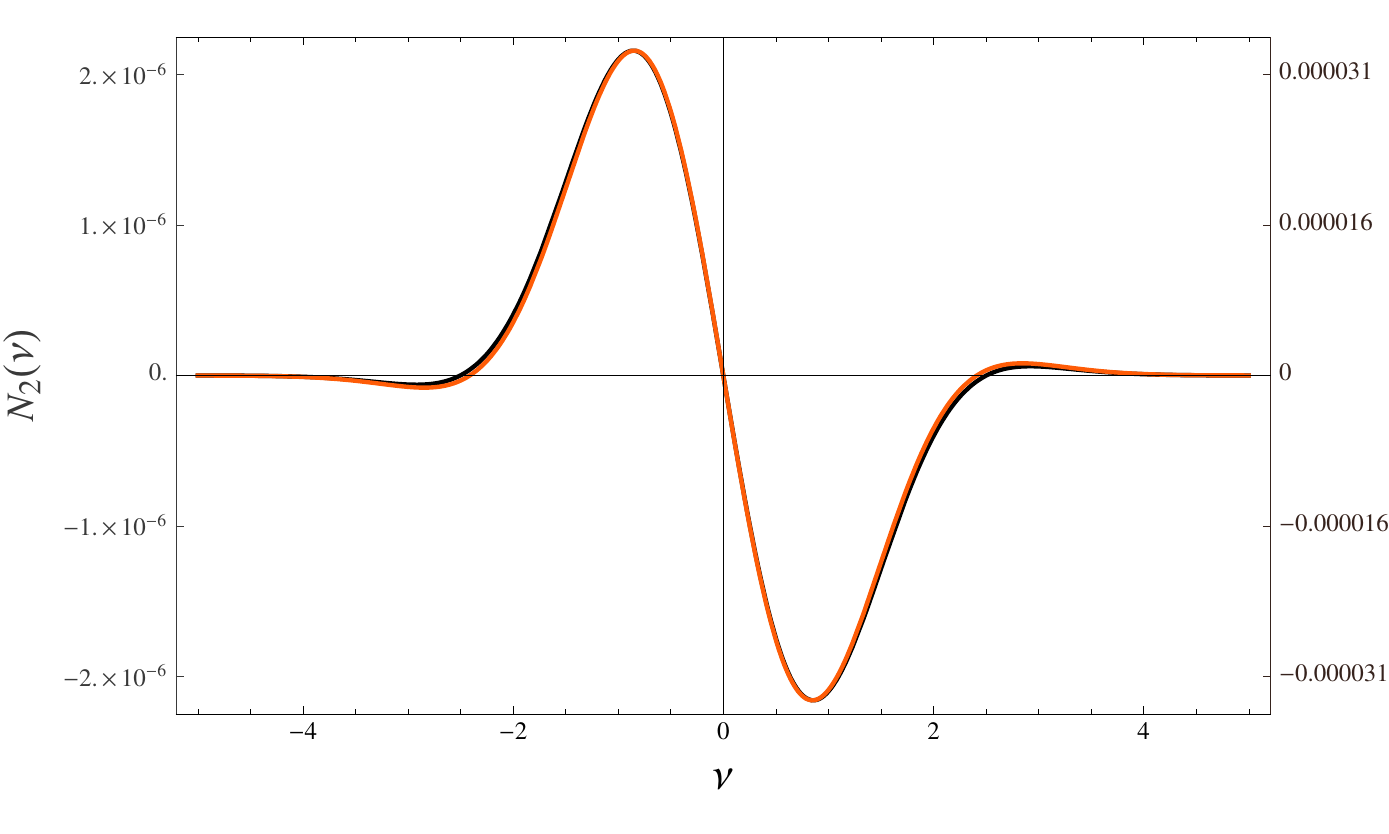}
\caption{The non-Gaussian correction to length of isocontours $\mathcal{N}_2(\nu)$ as a function of threshold $\nu$ for both quadratic and step potentials. Left axis for the quadratic potential and right for the step potential with $c=0.002$, $d=0.02M_{pl}$ and $\phi_s=15.66\;M_{pl}$ while the smoothing scale is $R=\tau_0/50$.}
\label{N2v}
\end{center}
\end{figure}

The second Minkowski functional is the length of isocontours (per unit volume) 
that is given by
\begin{eqnarray}
\mathcal{N}_2(\nu)=\frac{\sigma_1 e^{-\frac{\nu^2}{2}}}{2^{\frac{3}{2}}\sigma}\left( 1+ \sigma\frac{S_2}{6}H_3(\nu) +\sigma\frac{T_2}{2}H_1(\nu) +\mathcal{O}(\sigma^2) \right)
\label{eq:N2}
\end{eqnarray}
as a function of threshold $\nu$.
The first order correction in $\sigma$ to the length of isocontours
$\mathcal{N}_2$ is plotted as a function of threshold in Fig.~\ref{N2v}.
The shape of the curve is determined by the balance between $S_2$ and $T_2$,
with $T_2$ responsible for linear term, while $S_2$ for higher order cubic
behavior. In both models in Fig.~\ref{N2v} the $T_2$ contribution is dominant
leading to similar shape with single prominent maximum and minimum.
The effect of non-Gaussianity is again enhanced in the step potential 
model. 

\begin{figure}[htbp]
\begin{center}
\includegraphics[height=5.1cm]{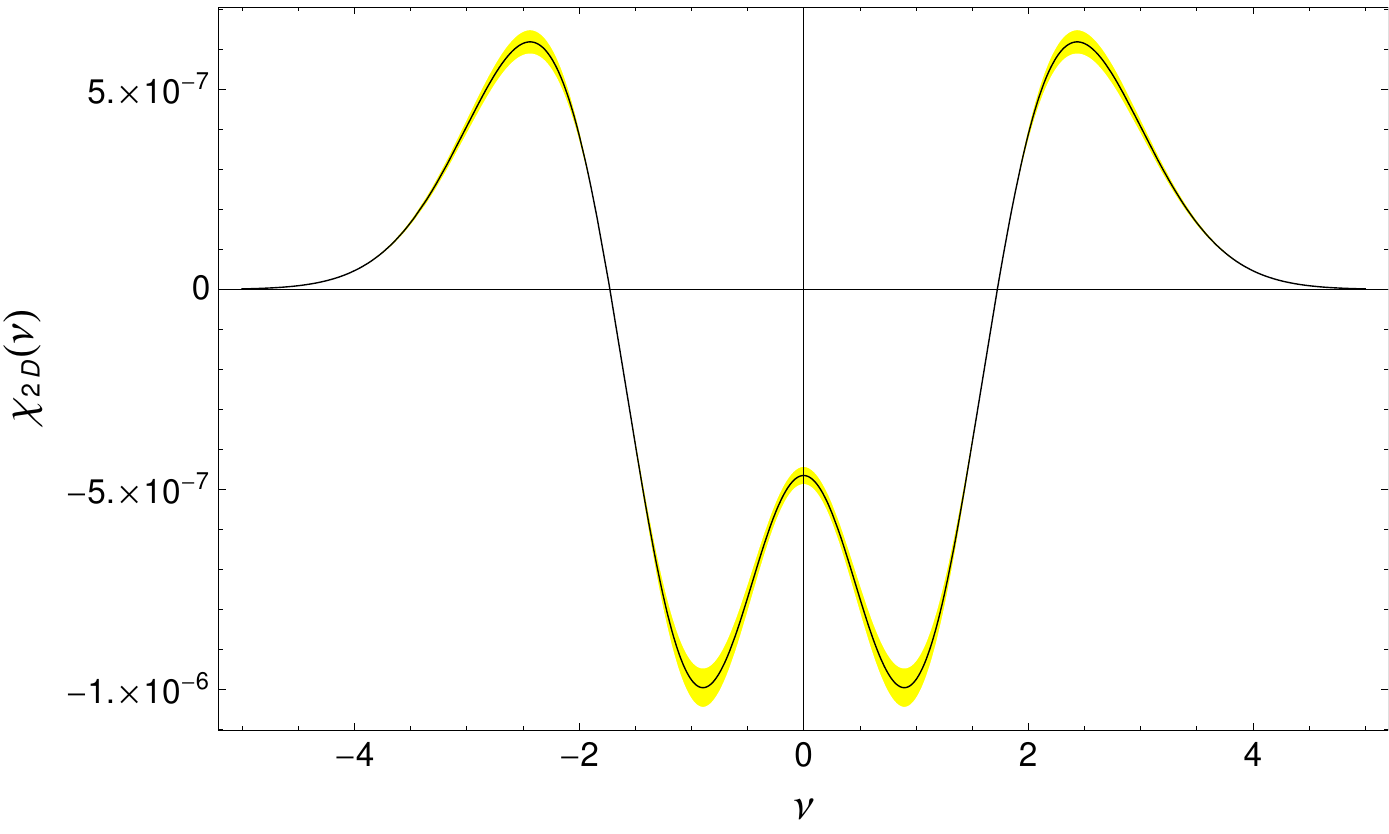}
\caption{The non-Gaussian correction to Euler characteristic $\chi_2(\nu)$ as a function of threshold $\nu$ for quadratic potential while the smoothing scale is $R=\tau_0/50$. Shaded yellow areas are numerical uncertainty in the value of the moments.}
\label{X2v}
\end{center}
\end{figure}

\begin{figure}[htbp]
\begin{center}
\includegraphics[height=5.2cm]{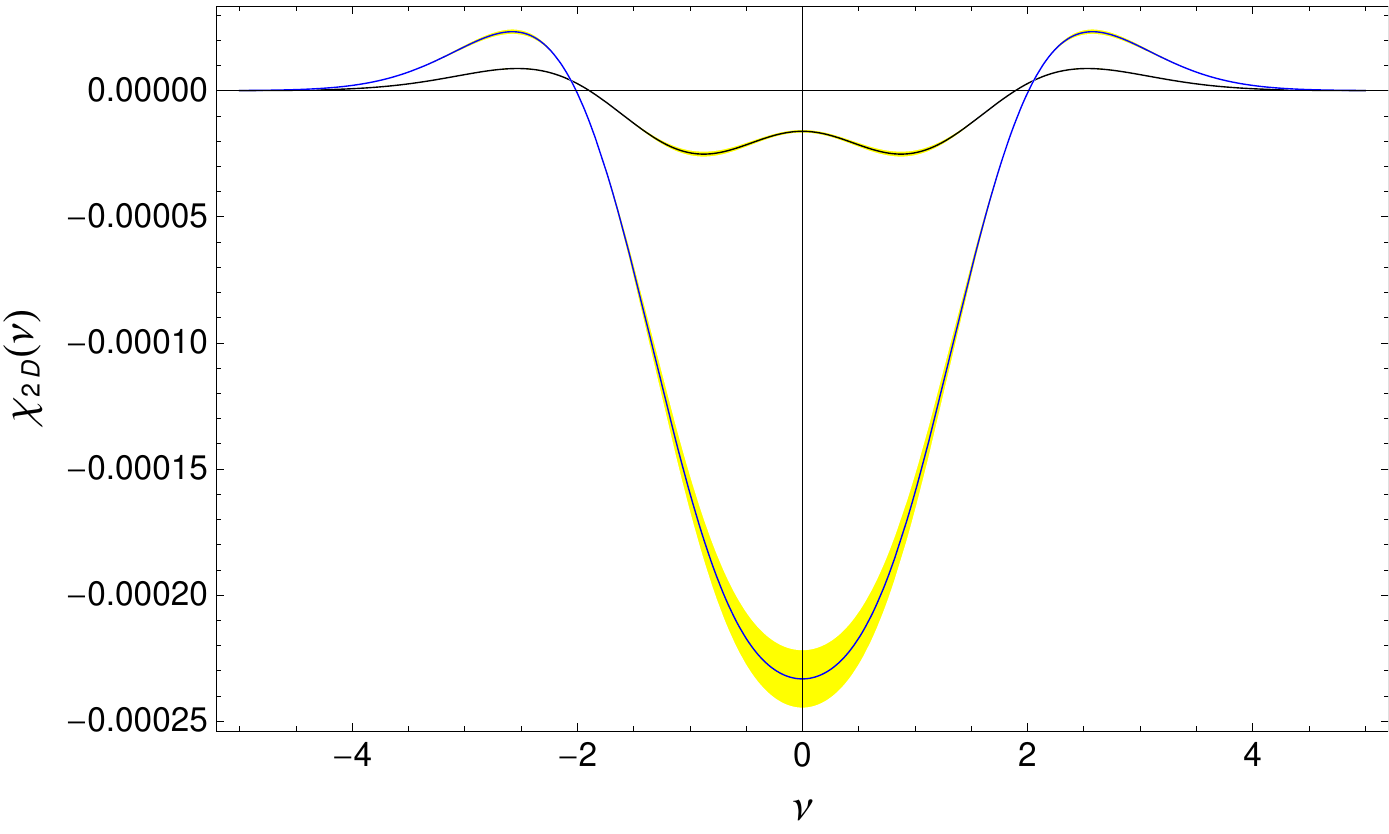}
\caption{The non-Gaussian correction to Euler characteristic $\chi_2(\nu)$ as a function of threshold $\nu$ for step potential with $c=0.002$, $d=0.02M_{pl}$ in black and $c=0.01$, $d=0.01M_{pl}$ in blue and $\phi_s=15.86\;M_{pl}$ for both, while the smoothing scale is $R=\tau_0/50$. Shaded yellow areas are numerical uncertainty in the value of the moments.}
\label{X2vs}
\end{center}
\end{figure}

The two-dimensional Euler characteristic or genus as function of threshold $\nu$ is given by
\begin{eqnarray}
\chi_\textsc{2}(\nu) &=& \left(\frac{\sigma_1}{\sqrt{2}\sigma} \right)^2 \frac{e^{-\nu^2/2}}{(2\pi)^{3/2}}
\Bigg[ H_1(\nu)\label{genus_2} \\
&~&+ \sigma\left(\frac{S_2}{6}H_4(\nu) - \frac{T_2}{2} H_2(\nu) - U_2 \right) +\mathcal{O}(\sigma^2) \Bigg]\notag~.
\end{eqnarray}
It involves all three studied moments,
$U_2$, $T_2$ and $S_2$, which contribute, respectively 
three lowest even order Hermite
contributions to the threshold dependence. We show example behavior
for quadratic and step potentials separately in 
Fig.~\ref{X2v} and Fig. \ref{X2vs}. 

\begin{figure}[htbp]
\begin{center}
\includegraphics[height=5.2cm]{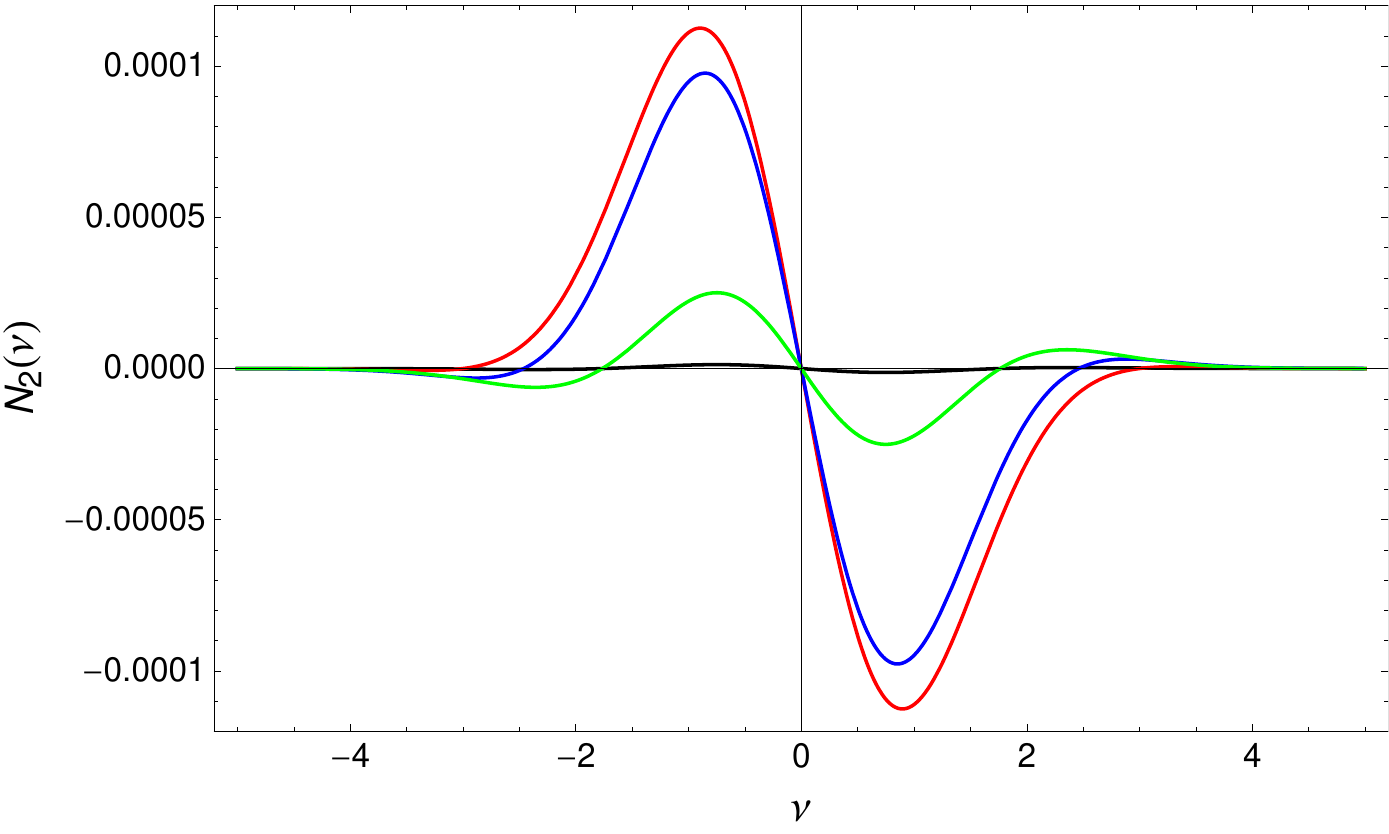}
\caption{The non-Gaussian correction to the length of isocontours 
$\mathcal{N}_2(\nu)$ as a function threshold $\nu$ for step potential with
$c=0.01$, $d=0.01 M_{pl}$ and $\phi_s=15.86 M_{pl}$ at different smoothing scales $R=\tau_0/2$ (in Black), $R=\tau_0/10$ (in Red), $R=\tau_0/50$ (in Blue) and $R=\tau_0/100$ (in Green).}
\label{N2R}
\end{center}
\end{figure}

\begin{figure}[htbp]
\begin{center}
\includegraphics[height=5.2cm]{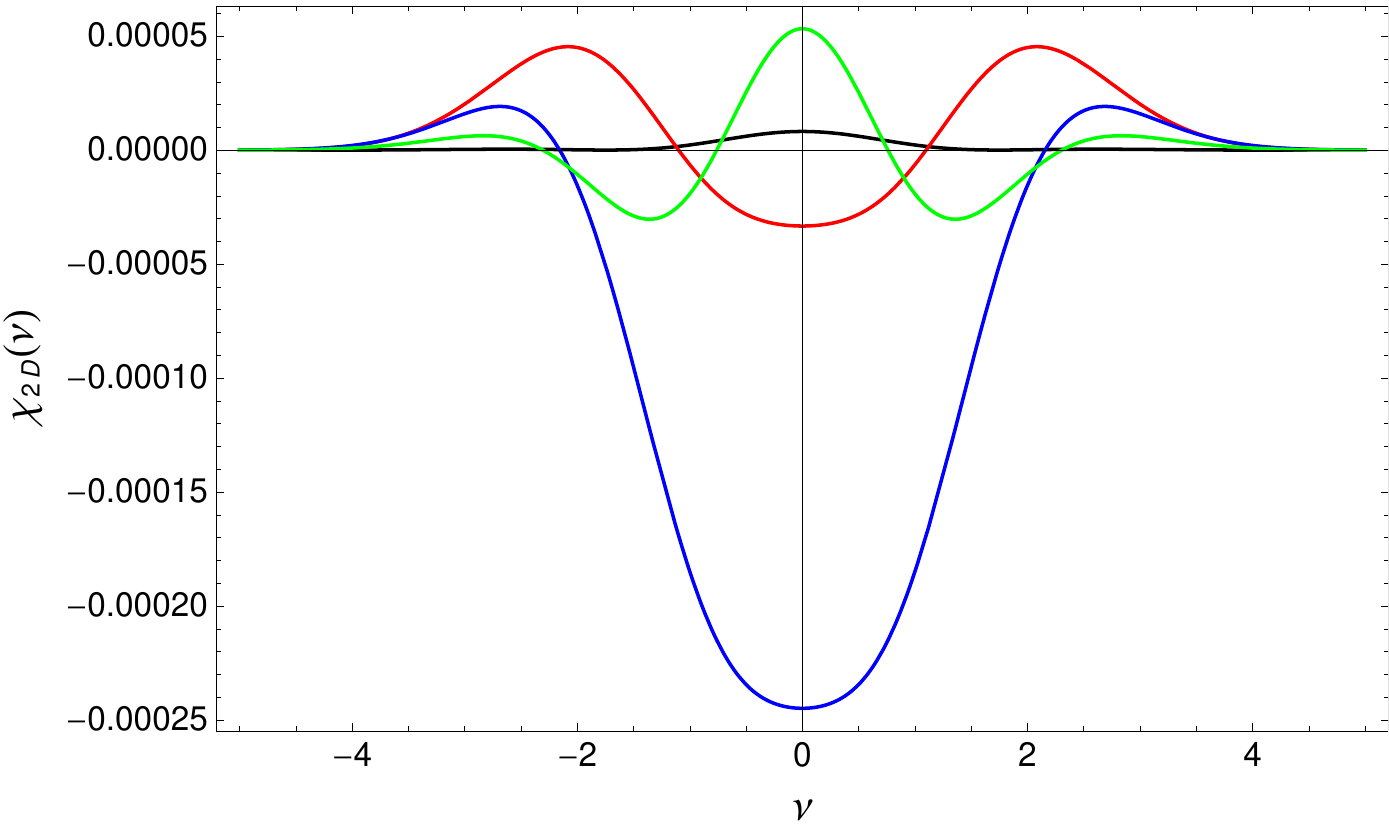}
\caption{The non-Gaussian correction to Euler characteristic $\chi_2(\nu)$ as a function threshold $\nu$ for step potential with $c=0.01$, $d=0.01 M_{pl}$ and $\phi_s=15.86 M_{pl}$ at different smoothing scales $R=\tau_0/2$ (in Black), $R=\tau_0/10$ (in Red), $R=\tau_0/50$ (in Blue) and $R=\tau_0/100$ (in Green).}
\label{X2R}
\end{center}
\end{figure}

The power in discriminating between different inflationary models comes from studying variation of MFs with smoothing scale, where
potentially the smoothing window can also be designed to maximize the detectability of the non-Gaussian features.
So, in Fig.~\ref{N2R}  we show the dependence of $\mathcal{N}_2$ statistics and
in Fig.~\ref{X2R} of Euler characteristic $\chi_2$ on $R$.

These figures show that when $R \le \tau_0/2$ (i.e to reaches into the step resonance region on 
the ultraviolet side of the position of the step in the inflaton potential),
the non-Gaussianity increases by a factor of ten for our choice of step
parameters. Subsequent behavior at the lesser smoothing scale is complex,
reflecting and correspondingly allowing one to reconstruct the
$R$ dependence of the moments shown in Fig.~\ref{Ms2}.

\begin{figure}[ht]
\begin{center}
\includegraphics[height=5.2cm]{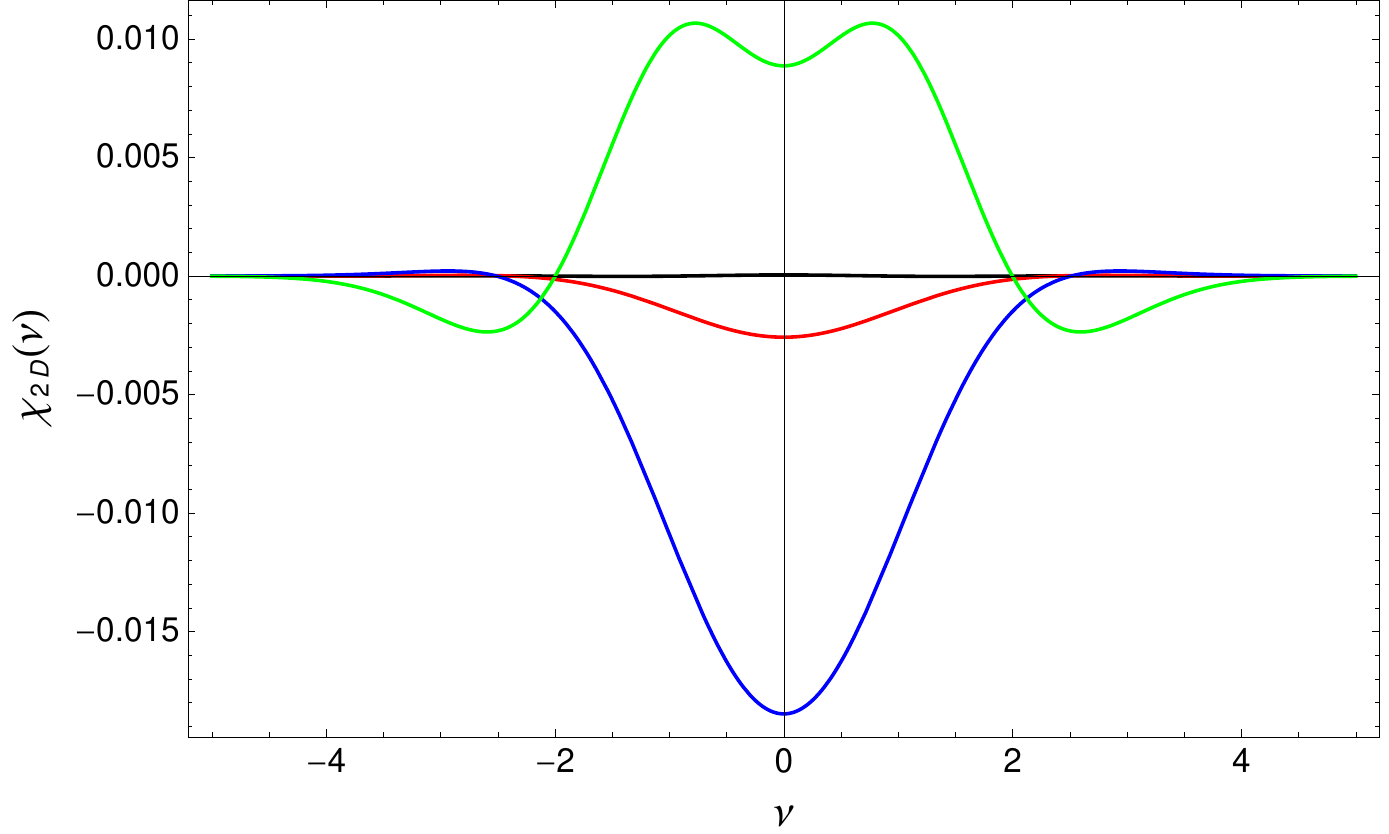}
\caption{The non-Gaussian correction to Euler characteristic $\chi_2(\nu)$ as a function threshold $\nu$ for step potential with $c=0.1$, $d=0.01 M_{pl}$ and $\phi_s=15.86 M_{pl}$ at different smoothing scales $R=\tau_0/2$ (in Black), $R=\tau_0/10$ (in Red), $R=\tau_0/50$ (in Blue) and $R=\tau_0/100$ (in Green).}
\label{X2S}
\end{center}
\end{figure}

The amplitude of non-Gaussianity in the step-potential models that we used as an example
for our technique is still lower than the uncertainty in measurements of Minkowski functionals
reported in \cite{planck3,Planck15} (which, for instance, is at the level of $\Delta \xi_{2D} \sim 0.01$ for the normalized Euler characteristic at full Planck resolution). Significant uncertainty in deducing primordial
non-Gaussianity from the data comes from non-Gaussian contributions from secondary effects - CMB lensing and
residuals after foreground subtraction. We can reach the level of non-Gaussianity in step-models 
that would easily be distinguishable in the data if we raise the height of step parameter to $c=0.1$.
In Fig.~\ref{X2S} the non-Gaussian correction to the normalized Euler characteristic is plotted for this case
showing corrections of  $0.01$ magnitude. However, at such large steps, the distortion of the power spectrum
is also significant, and the joint analysis of non-Gaussian Minkowski functionals and power spectrum is
needed to see which effect is more constraining. 


In Appendix~\ref{Mofvf} we present an alternative approach to studying Minkowski
functionals as the functions of filling factor $f_{V_2}$ rather than
of the field threshold value $\nu$ \cite{Pogosyan2}.
Such approach is needed when the
variance of the field cannot be determined with sufficient precision
making difficult to find $\nu$, which involves scaling of the field 
by the variance. This is a frequent situation with the large-scale structure
data for galaxy distribution, where coverage may be biased toward
overdense regions. Large uncertainty in the variance is usually not an issue
for CMB all-sky maps; nevertheless, the technique may still be useful.

In summary of this section, we have shown how Minkowski functionals for CMB maps are 
computed from the first principles in single field inflationary models, with the step
potential model used as a nontrivial example.
It is found that the magnitude of Minkowski functionals is significantly
larger for step potential when compared with slow-roll models of inflation.
The shape of the MFs as functions of threshold
also differs between different models of inflation.
It is also noted, as in Fig. \ref{X2S}, that one can find parameter values where these MFs can be constrained by Planck analysis at the level of \cite{planck3,Planck15}.
It is demonstrated that MFs change in model dependent way
with the smoothing cutoff $R$. Thus, we can can conclude that the Minkowski
functionals are an efficient tool to link non-Gaussianity in two-dimensional
CMB data to the nonlinear processes on inflation in models with 
features in the inflaton potential.


\section{Results and Discussion}
The goal in this work was to develop a link between the two-dimensional Minkowski functionals
for temperature anisotropy maps and different cosmic inflationary models. Thus, we developed
a robust bridge between the early Universe inflationary models
and the observable 2D geometrical characteristics
of the initial field of scalar adiabatic cosmological perturbations.
The link to non-Gaussian features in such observables as Minkowski functionals,
extrema counts, and skeleton properties is provided by studying the
higher order moments of the perturbation field and its derivatives in
configuration space \cite{Matsubara1,Bardeen3,PGP09,PPG11,Pogosyan1,Pogosyan2}.

We have investigated from ab initio the third-order configuration
space moments for 2D temperature maps that give first non-Gaussian corrections.
To calculate these moments we have to calculate the
three-point bispectrum in momentum space and then integrate over the three
momenta using spherical harmonics for 2D maps of the sky.
We calculated the moments using the transfer functions generated by CAMB 
\cite{Lewis12}, but using instant recombination approximation, suitable for
$\ell < 100$. Monopole subtraction was performed in the momentum space,
and this assured the infrared convergence of the calculations.

We applied these 2D moments to the study of the CMB temperature fluctuations 
map that is the most direct probe of non-Gaussianity.
The step potential model that breaks slow-roll conditions shows interesting dependence of moments on 
smoothing cutoff $R$ and their values increase by a factor of $100$ for some values of $R$.
For this model, we have larger parameter space of $c$, $d$ and $\phi_s$
for which we have to find optimal $R$ and other parameters for detection at different scales. Thus, these Minkowski functionals give distinctly different
signature for the slow-roll models and for the models that break slow-roll conditions as the step potential model.

\section{Acknowledgments}
This research has been supported by Natural Sciences 
and Engineering Research Council (NSERC) of Canada 
Discovery Grant. The computations were performed on the 
SciNet HPC Consortium. SciNet is funded by: the Canada 
Foundation for Innovation under the auspices of Compute 
Canada; the Government of Ontario; Ontario Research 
Fund - Research Excellence; and the University of Toronto\cite{SciNet}.

\bibliographystyle{unsrt} 
\bibliography{ThesisRefs}

\appendix \label{momo}
\section{General Formula 2D Moments}
The calculation of moments in 2D consists of three integration as can be seen in Eq. \ref{tilde_x3}. One of the integrations is taken out with help of the delta function and integral then takes the following general form 
\begin{eqnarray}
\int \frac{dk_1^3}{(2\pi)^3}\frac{dk_2^3}{(2\pi)^3} f(\bold{k_1}) B(k_1,k_2,\bold{k_1}\cdot\bold{k_2})g(\bold{k_2})\label{GF}
\end{eqnarray}
in terms of the bispectrum $B(k_1,k_2,\bold{k_1}\cdot\bold{k_2})$. We can in general decompose the functions $f(\bold{k_1})$, $g(\bold{k_2})$ and $B(k_1,k_2,\bold{k_1}\cdot\bold{k_2})$ in the above integral into spherical harmonics as follows
\begin{eqnarray}
f(\bold{k_1}) &=& \sum_{lm} f_{lm}(k_1)Y_{lm}(\hat{k}_1), \\
g(\bold{k_2}) &=& \sum_{lm} f_{lm}(k_2)Y_{lm}(\hat{k}_2),\notag \\
B(k_1,k_2,\bold{k_1}\cdot\bold{k_2}) &=& 4\pi \sum_{lm} B_l(k_1,k_2) Y_{lm}(\hat{k}_1)Y_{lm}^*(\hat{k}_2).\notag
\end{eqnarray}
Using the above relations we find a general expression for the integral in Eq.~\ref{GF} as
\begin{eqnarray}
&\Rightarrow&\int \frac{dk_1^3}{(2\pi)^3}\frac{dk_2^3}{(2\pi)^3} f(\bold{k_1}) B(k_1,k_2,\bold{k_1}\cdot\bold{k_2})g(\bold{k_2})\notag \\
&=&\int \frac{k_1^2dk_1}{(2\pi)^3}\frac{k_2^2 dk_2}{(2\pi)^3} 4\pi \sum_{lm}\sum_{l'm'}\sum_{l''m''} f_{l'm'}(k_1) B_l(k_1,k_2) \notag \\
&~&g_{l''m''}(k_2) \int d\Omega_1 d\Omega_2 Y_{l'm'}(k_1)Y_{lm}(k_1)   Y_{lm}(k_2)Y_{lm}^*(k_2)\notag \\
&=&\int \frac{k_1^2dk_1}{(2\pi)^3}\frac{k_2^2 dk_2}{(2\pi)^3} 4\pi \sum_{lm}\sum_{i'm'}\sum_{l''m''} f_{l'm'}(k_1) B_l(k_1,k_2)\notag \\
&~&\quad g_{l''m''}(k_2) (-1)^m\delta_{ll'}\delta_{m,-m'}~ \delta_{ll''}\delta_{mm''}\notag \\
&=&\int \frac{k_1^2dk_1}{(2\pi)^3}\frac{k_2^2 dk_2}{2\pi^2} \sum_{lm}(-1)^m f_{l,-m}(k_1) B_l(k_1,k_2)g_{lm}(k_2)\notag \\
\label{GSol}
\end{eqnarray}
Thus, we can find the solution to cubic moments by expanding $f(k_1)$, $g(k_2)$ and $B(k_1,k_2,\bold{k_1}\cdot\bold{k_2})$ in multipoles and insert them in Eq.~\ref{GSol} to get the result for different moments.

\section{Calculation of Cubic Moments}
The temperature maps have a monopole element(the average temperature) that needs to be subtracted to get the physical results, i.e, the temperature fluctuations,	
\begin{eqnarray}
&~&\left< \widetilde{\zeta}^3 \right> = \int d\Omega_{ps} \Big( e^{i(\bold{k_1}+\bold{k_2} + \bold{k_3})\cdot\bold{\hat{n}}\tau_0}\notag \\&~& - j_0(k_1\tau_0)j_0(k_2\tau_0)j_0(k_3\tau_0) + j_0(k_1\tau_0)j_0(k_2\tau_0)e^{i\bold{k_3}\cdot\bold{\hat{n}}\tau_0} \notag \\
&~&+ j_0(k_2\tau_0)j_0(k_3\tau_0)e^{i\bold{k_1}\cdot\bold{\hat{n}}\tau_0}  + j_0(k_1\tau_0)j_0(k_3\tau_0)e^{i\bold{k_2}\cdot\bold{\hat{n}}\tau_0} \notag \\
&~& -\Big. j_0(k_1\tau_0)e^{i(\bold{k_2}+\bold{k_3})\cdot\bold{\hat{n}}\tau_0} - j_0(k_2\tau_0)e^{i(\bold{k_3}+\bold{k_1})\cdot\bold{\hat{n}}\tau_0} \notag \\ 
&~& - j_0(k_3\tau_0)e^{i(\bold{k_1}+\bold{k_2})\cdot\bold{\hat{n}}\tau_0}  \Big)\notag.
\end{eqnarray}
We calculated term by term with first term $e^{i(\bold{k_1}+\bold{k_2}+\bold{k_3})\cdot\bold{\hat{n}}\tau_0}$ being just the three-point function calculation. Thus, we get the monopole correction for the cubic moment $\left< \widetilde{\zeta}^3 \right> $ by summing all the above five terms to get
\begin{eqnarray}
\left< \widetilde{\zeta}^3 \right> &=&\int d\Omega_{ps} \left(1+2j_0(k_1\tau_0)j_0(k_2\tau_0)j_0(|\bold{k}_3|\tau_0)\right. \notag \\ &~& \left.- \frac{3}{2} \left( j^2_0(k_1\tau_0) +j^2_0(k_2\tau_0)  \right)  \right).
\end{eqnarray}
In the above expression we see that in the limit $k_1\rightarrow 0$ and $k_2\rightarrow 0$, the $\left< \widetilde{\zeta}^3 \right>$ moment vanishes. This shows that the monopole contribution has been successfully eliminated from the $S_3$ moment.

Next, we calculate the second moment $\left< \widetilde{\zeta}^2 \Delta\widetilde{\zeta} \right>$ with the monopole eliminated that is given below
\begin{eqnarray}
&~&\left< \widetilde{\zeta}^2\Delta\widetilde{\zeta} \right> 
=-\int d\Omega_{ps} k_{3\perp}^2 \tau_0^2\Big( 1 -  j_0(k_1\tau_0)e^{-i\bold{k_1}\cdot\bold{\hat{n}}\tau_0}\notag \\  
&~& - j_0(k_2\tau_0)e^{-i\bold{k_2}\cdot\bold{\hat{n}}\tau_0}+j_0(k_1\tau_0)j_0(|\bold{k}_1+\bold{k}_3|\tau_0)e^{i\bold{k_3}\cdot\bold{\hat{n}}\tau_0} \Big)\notag,
\end{eqnarray}
where we have used plane perpendicular approximation $\Delta_{\bold{\hat{n}}}e^{i\bold{k_3}\cdot\bold{\hat{n}}\tau_0}=-k_{3\perp}^2 \tau_0^2 e^{i\bold{k_3}\cdot\bold{\hat{n}}\tau_0}$ that is evaluated on the sphere at North Pole. In the calculation of this moment we consider a symmetric bispectrum in momenta($k_1$, $k_2$, $k_3$) with no special choice for the delta function. Thus, the $\left< \widetilde{\zeta}^2\Delta\widetilde{\zeta} \right>$ moment takes the following form
\begin{eqnarray}
\left< \widetilde{\zeta}^2 \Delta\widetilde{\zeta} \right>
&=& - \int d\Omega_{ps} \frac{2}{3}k_2^2 \tau_0^2 \Bigg( 1+ j_0(|\bold{k}_3|\tau_0) \notag
\\&~&\times j_0(k_1\tau_0)\left( j_0(k_2\tau_0)+j_2(k_2\tau_0) \right) \notag
\\ &~& -2\left(j_0(k_1\tau_0) +j_2(k_1\tau_0)P_2(\cos\theta) \right)\Bigg).~~\label{tilde_x2y} 
\end{eqnarray}
The above result can also be obtained from the general solutions of the three-point function in Eq.~\ref{GSol}.

Lastly, we evaluate the third moment 
$\left\langle (\nabla \widetilde{\zeta})^2 \Delta\widetilde{\zeta} \right\rangle$ that is independent of the monopole due to the derivatives. This moment is finite in the infrared regime; thus, it is more sensitive to the ultraviolet cutoff, which is why we use a smooth Gaussian cutoff scheme. Moreover, to calculate this moment on the 2D sphere we use the following relationship:
\begin{eqnarray}
\left\langle (\nabla \widetilde{\zeta})^2 \Delta\widetilde{\zeta} \right\rangle
&=& -\left< \zeta \left(\Delta\zeta \right)^2\right> + \frac{1}{2}\left< \zeta^2\Delta^2\zeta \right>\label{dz2dz}
\end{eqnarray}
So the third moment $\left< \zeta(\tau_0\bold{\hat{n}})\left(\Delta\zeta(\tau_0\bold{\hat{n}})\right)^2\right>$ is evaluated to be
\begin{eqnarray}
\left< (\nabla \widetilde{\zeta})^2 \Delta\widetilde{\zeta} \right>
&=&  \frac{4}{3} \int d\Omega_{ps}  \tau_0^4\left(  \frac{k_{2}^4+k_3^4}{10}\right. \notag \\  &~& \left. -\frac{k_{2}^2 k_{3}^2}{3}\left(1+ \frac{P_2(\cos\theta)}{5}\right) \right).\quad \label{tilde_q2y}
\end{eqnarray}
This moment, containing only derivatives, is independent of the monopole term.

\section{Minkowski functionals as functions of the filling factor}
\label{Mofvf}

When direct determination of the second moments of the field $\sigma$ and
$\sigma_1$ is not sufficiently accurate, the
Minkowski functionals can be analyzed as functions of the filling factor
$f_{V_2}$, which can often be measured even on incomplete data.
The procedure is the following. The filling factor and the rest of the
MFs are simultaneously 
measured for a set of real field thresholds 
(not scaled by the variance $\sigma$), and then $\mathcal{N}_2$ and
$\chi_2$ are expressed with respect to $f_{V_2}$. 

\begin{figure}[t]
\begin{center}
\includegraphics[height=5.2cm]{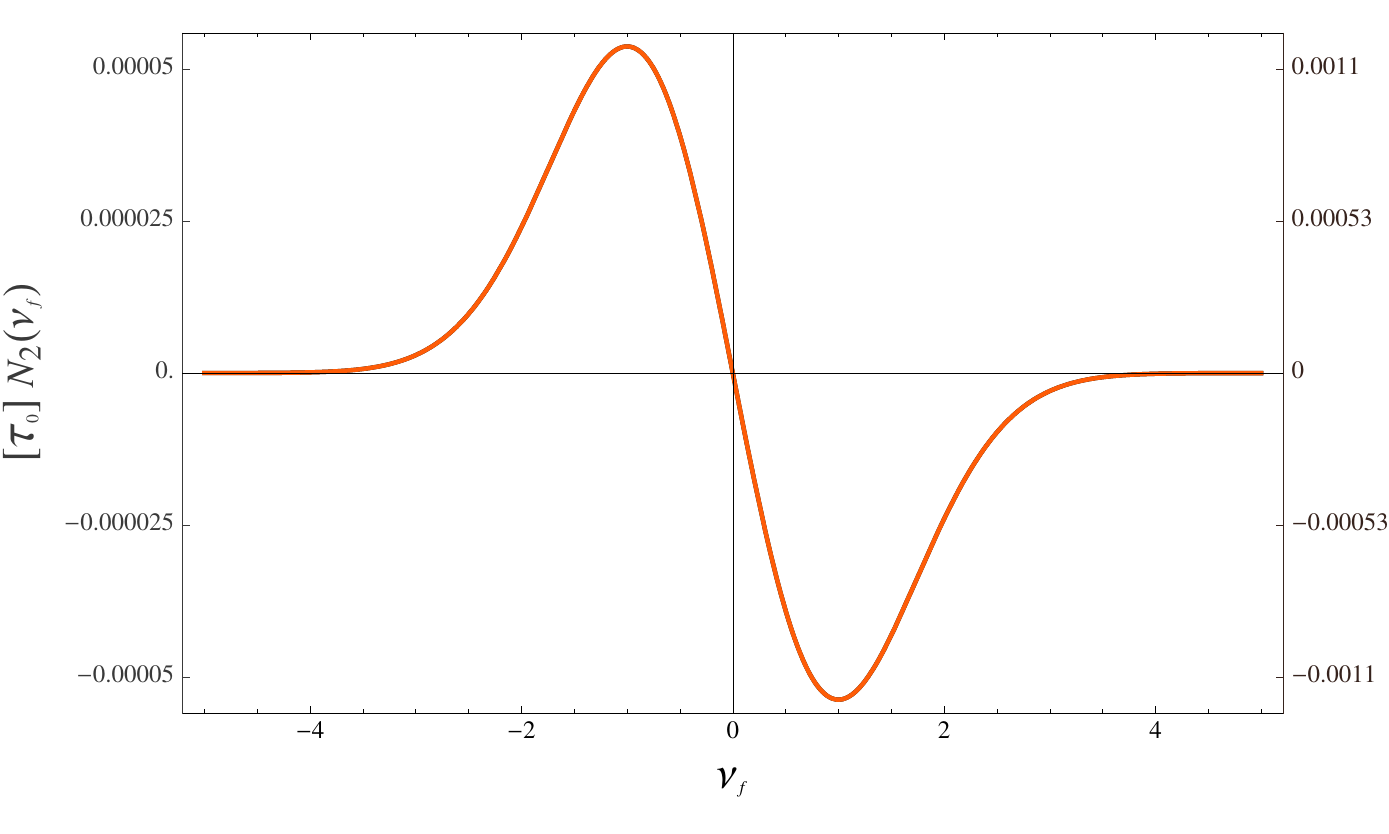}
\caption{The non-Gaussian correction to length of isocontours $\mathcal{N}_2(\nu_f)$ as a function of filling factor $\nu_f$ for both quadratic and step potentials. Left axis for the quadratic potential and right for the step potential with $c=0.002$, $d=0.02M_{pl}$ and $\phi_s=15.76\;M_{pl}$ while the smoothing scale is $R=\tau_0/50$.}
\label{N2f}
\end{center}
\end{figure}

\begin{figure}[t]
\begin{center}
\includegraphics[height=4.8cm]{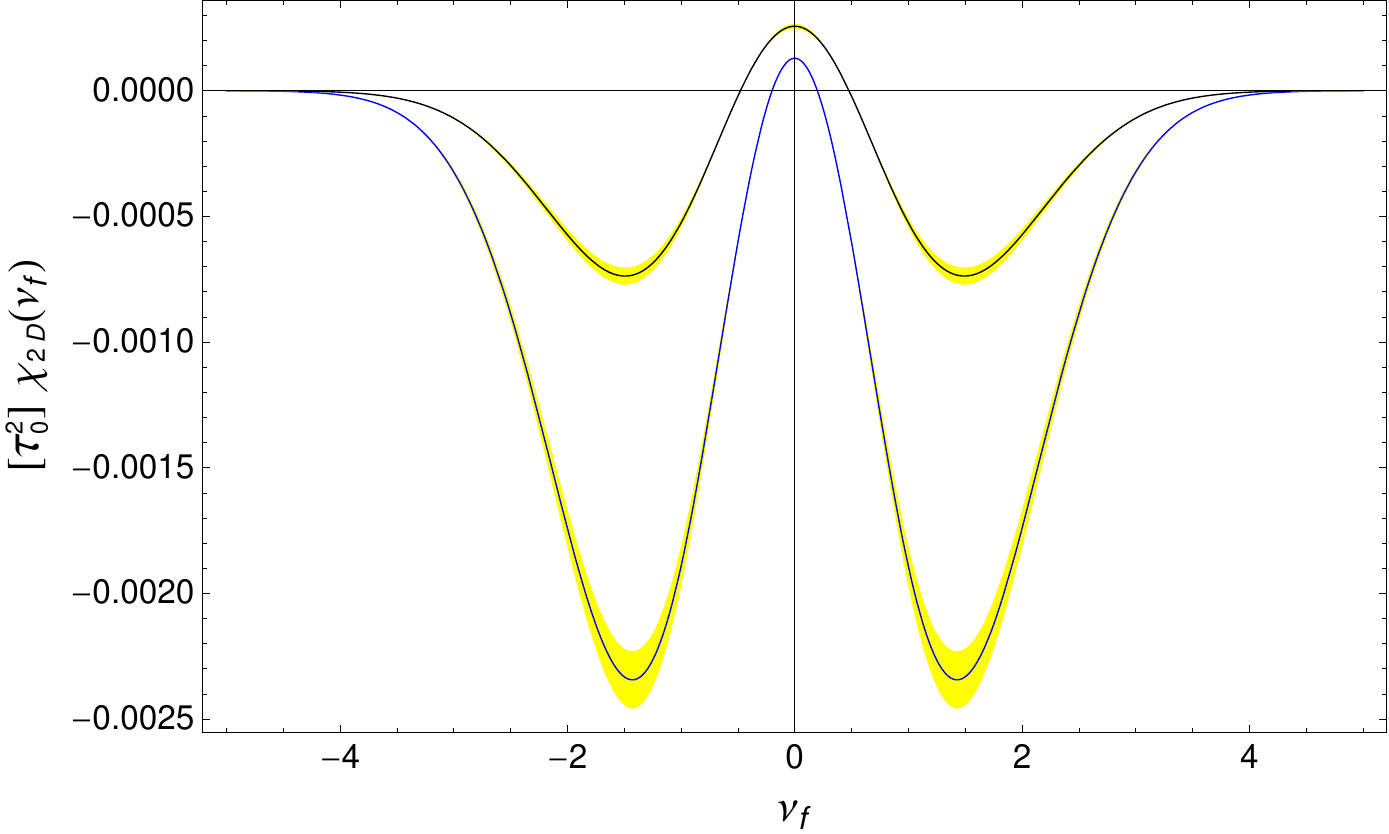}
\caption{The non-Gaussian correction to Euler characteristic $\chi_2(\nu_f)$ as a function of $\nu_f$ for step potential with $c=0.002$, $d=0.02M_{pl}$ in black and $c=0.01$, $d=0.01M_{pl}$ in red and $\phi_s=15.86\;M_{pl}$ for both, while the smoothing scale is $R=\tau_0/50$. Shaded yellow areas are numerical uncertainty in the value of the moments.}
\label{X2vf}
\end{center}
\end{figure}

\begin{figure}[t]
\begin{center}
\includegraphics[height=4.8cm]{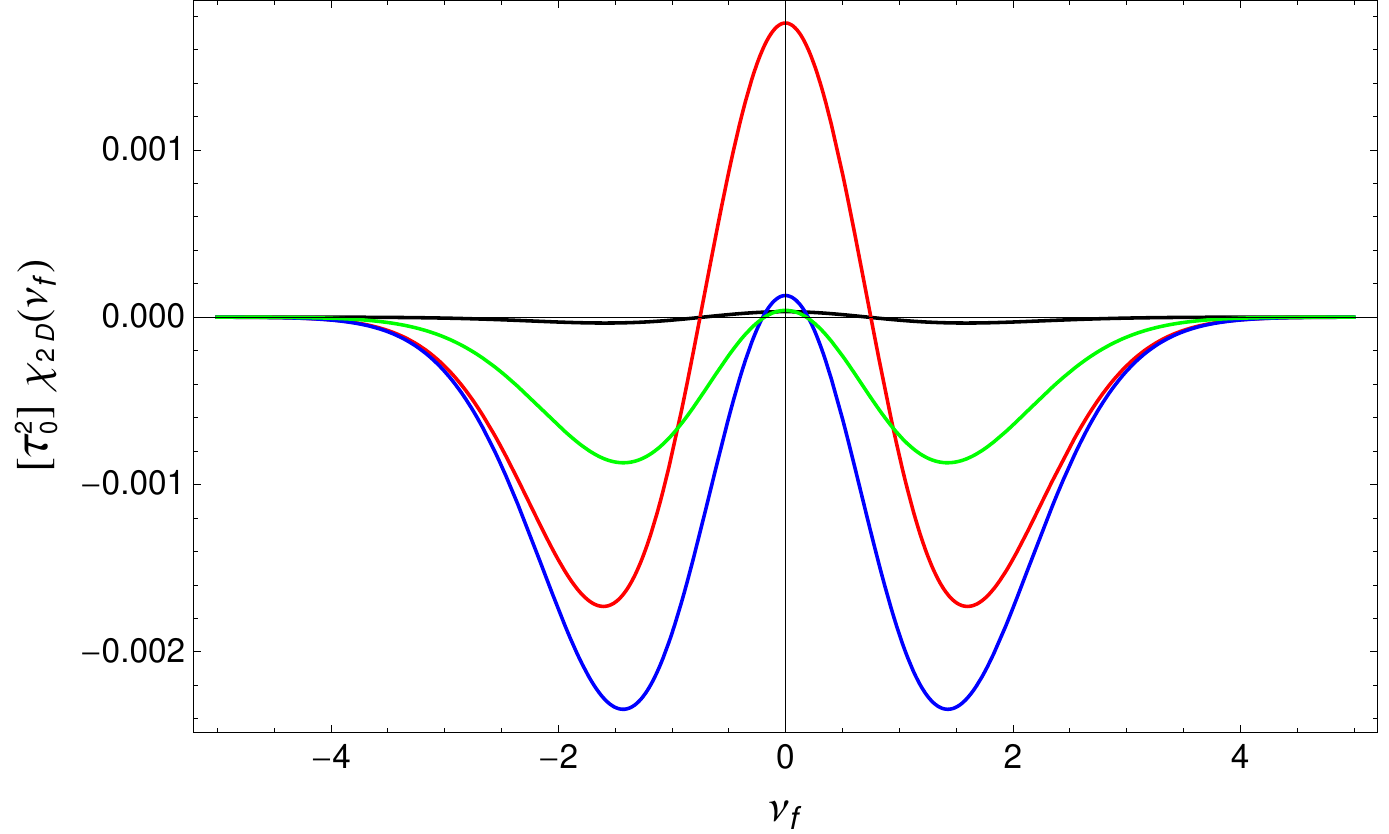}
\caption{The non-Gaussian correction to Euler characteristic $\chi_2(\nu_f)$ as a function of filling factor $\nu_f$ for step potentials with $c=0.01$, $d=0.01M_{pl}$ and $\phi_s=15.86\;M_{pl}$ at different smoothing scales $R=\tau_0/2$ (in Black), $R=\tau_0/10$ (in Red), $R=\tau_0/50$ (in Blue) and $R=\tau_0/100$ (in Green).}
\label{X2vfs}
\end{center}
\end{figure}

Mathematically, we introduce the effective $\nu_f$ that represents the filling factor
via the definition $\frac{1}{2} \mathrm{erfc}(\nu_f/\sqrt{2}) \equiv f_{V_2}$.
After inverting relation~(\ref{eq:fv}), we find to the first non-Gaussian order
that
\begin{equation}
\nu = \nu_f + \sigma \frac{S_2}{6} H_2(\nu_f) +\mathcal{O}(\sigma^2) ~.
\end{equation}
After substituting this relation into Eqs.~(\ref{eq:N2}) and (\ref{genus_2})
we see that
\begin{eqnarray}
\mathcal{N}_2(\nu_f) \approx 
\frac{\sigma_1 e^{-\frac{\nu_f^2}{2}}}{2^{\frac{3}{2}}\sigma}
\left( 1 +\sigma\left(\frac{T_2}{2}-S_2\right)H_1(\nu_f) \right).
\end{eqnarray}
and 
\begin{eqnarray}
\chi_\textsc{2}(\nu_f) &\approx& \left(\frac{\sigma_1}{\sqrt{2}\sigma} \right)^2 \frac{e^{-\nu_f^2/2}}{(2\pi)^{3/2}}
\Bigg[ H_1(\nu_f)\label{genus_2f} \\ 
&~& -\sigma\left(S_2 +\frac{U_2}{2} \right) -\sigma\left( S_2 - \frac{T_2}{2} \right)H_2(\nu_f) \Bigg]\notag
\end{eqnarray}
Thus, fitting the length of isocontours $\mathcal{N}_2(\nu_f)$ and Euler
characteristic $\chi_2(\nu_f)$ statistics to the data
as an expansion into $H_0$, $H_1$ and $H_2$ Hermite polynomials we determine
$\sigma_1/\sigma$ from the Gaussian terms and two combinations,
$\sigma ( S_2 - T_2/2) $ and $\sigma (S_2 + U_2/2)$ from the non-Gaussian 
corrections.
Figure~\ref{N2f} shows the non-Gaussian part of isocontour length $N_2$ as function of $\nu_f$ that has the universal shape. The model dependent information
is in the amplitude proportional to $\sigma \left(\frac{T_2}{2}-S_2\right)$
and its dependence on the smoothing scale $R$.

The Euler characteristic $\chi_2(\nu_f)$ gives additionally the access to $\sigma (S_2 + U/2)$
and exhibits
functional form changes as the underlying model and smoothing scale changes.
This can be viewed in Figs.~\ref{X2vf} and \ref{X2vfs}.
The constant contribution of 
$\sigma(S_2 +U_2/2)$ to non-Gaussian terms results in a shift of
oscillatory curves with respect to the $x$-axis whereas the amplitude of oscillations
is set by $\sigma (S_2-T_2/2)$.

\end{document}